%% file: draftv13.tex
\documentclass[aps,prx,superscriptaddress,twocolumn,tightenlines,showpacs,floatfix,amsmath,amssymb,pdftex]{revtex4-1}
\usepackage[english]{babel}
\usepackage{mathtools}
\usepackage{amsmath}
\usepackage{amssymb}
\usepackage{braket}
\usepackage[normalem]{ulem}
\usepackage[left=2.5cm,right=2.5cm,top=2.5cm,bottom=2.5cm]{geometry}
\usepackage[pdftex,dvipsnames]{xcolor}
\usepackage{todonotes}
\usepackage{longtable}
\usepackage{theorem}
\usepackage[bookmarks=true,bookmarksnumbered=false, hyperindex=true,bookmarksopen=true,hyperfigures=true, colorlinks=true,linkcolor=webred,citecolor=webgreen,urlcolor=webblue,breaklinks]{hyperref}
\definecolor{webgreen}{rgb}{0, 0.5, 0}
\definecolor{webblue}{rgb}{0, 0, 0.5}
\definecolor{webred}{rgb}{0.5, 0, 0}
\usepackage{url}
\usepackage{setspace}
\usepackage{amsmath,amssymb,amscd}
\usepackage{amsfonts}
\usepackage{color}
\usepackage{graphicx}

\def\ben{\begin{equation}}
\def\een{\end{equation}}

\let\a=\alpha    
   
  \let\n=\nu   \let\r=v
\let\s=\sigma

 \let\G=\Gamma \let\D=\Delta  \let\L=\Lambda
   
 \let\W=\Omega

\def\be{\begin{equation}}
\def\ee{\end{equation}}
\def\ba{\begin{align}}
\def\ea{\end{align}}

\def\dalemb#1#2{{\vbox{\hrule height .#2pt
       \hbox{\vrule width.#2pt height#1pt \kern#1pt
               \vrule width.#2pt}
       \hrule height.#2pt}}}

\newcommand{\bea}{\begin{eqnarray}}
\newcommand{\eea}{\end{eqnarray}}

\newcommand{\Tr}{{\rm Tr} }

\theoremstyle{plain}
\newtheorem{thm}{Theorem}[section]

\newtheorem{prop}[thm]{Proposition}
\theoremstyle{definition}

\theoremstyle{remark}

\begin{document}
\title{Topological classification of crystalline insulators through band structure combinatorics}
\date{\today}

\author{Jorrit Kruthoff}
\affiliation{Institute of Physics, Universiteit van Amsterdam,
Science Park 904, 1090 GL Amsterdam, The Netherlands}
\author{Jan de Boer}
\affiliation{Institute of Physics, Universiteit van Amsterdam,
Science Park 904, 1090 GL Amsterdam, The Netherlands}
\author{Jasper van Wezel}
\affiliation{Institute of Physics, Universiteit van Amsterdam,
Science Park 904, 1090 GL Amsterdam, The Netherlands}
\author{Charles L. Kane}
\affiliation{Department of Physics and Astronomy, University of Pennsylvania, Philadelphia, PA 19104-6323, USA}
\author{Robert-Jan Slager}
%\affiliation{Instituut-Lorentz for Theoretical Physics, Universiteit Leiden,
%PO Box 9506, NL-2300 RA Leiden, The Netherlands}
\affiliation{Max-Planck-Institut f\"ur Physik komplexer Systeme, 01187 Dresden, Germany}

\begin{abstract}
We present a method for efficiently enumerating all allowed, topologically distinct, electronic band structures within a given crystal structure. The algorithm applies to crystals with broken time-reversal, particle-hole, and chiral symmetries in any dimension. The presented results match the mathematical structure underlying the topological classification of these crystals in terms of $K$-theory, and therefore elucidate this abstract mathematical framework from a simple combinatorial perspective. Using a straightforward counting procedure, we classify the allowed topological phases in any possible two-dimensional crystal in class $A$. We also show how the same procedure can be used to classify the allowed phases for any three-dimensional space group. Employing these classifications, we study transitions between topological phases within class $A$ that are driven by band inversions at high symmetry points in the first Brillouin zone. This enables us to list all possible types of phase transitions within a given crystal structure, and identify whether or not they give rise to intermediate Weyl semimetallic phases.
\end{abstract}

\maketitle
\section{Introduction}
Over the past two decades, topological order has been established as an organising principle in the classification of matter, alongside the traditional symmetry-based approach. The discovery of the quantum spin Hall effect and topological insulators \cite{Hasan2010, Qi2011} however, made it apparent that symmetry cannot be ignored within the topological classification of even free fermion matter. The topological identities of these systems in fact depend on, and are protected by, the presence of an underlying symmetry \cite{Xie2013}. While the theoretical description of symmetry protected topological phases (SPTs) is connected to exotic field theories and branches of mathematics \cite{Wen2013}, their interplay between topology and symmetry also has direct and prominent experimental consequences. These include the presence of single protected edge states that circumvent the usual fermion doubling theorem, and the possibility of excitations having fractional charges and statistics \cite{Beenakker2013, Elliot2015}. 

The tenfold periodic table has been a cornerstone in the description of the connection between topology and symmetry~\cite{Kitaev2009, Ryu2010}. It specifies the number of topologically distinct ground states that are possible in free fermion systems in any number of dimensions, if their behaviour under time-reversal symmetry, particle-hole symmetry, and chiral symmetry is given~\cite{Altland1997}. The combinations of discrete symmetries on which the ten classes in the table are based, do not include any spatial symmetries. Materials in nature however, are made up out of atoms which are often positioned in a periodic crystal structure containing crystal symmetries. Indeed, the very existence of periodic band structures for electrons is a consequence of the breaking of translation symmetry by an atomic lattice. It is well-known that within time-reversal symmetric topological insulators, the discrete translational symmetries surviving within the atomic lattice lead to the definition of weak invariants in three dimensions, which need to be used in addition to the tenfold periodic table to get a full classification of the topological state~\cite{Fu2007, Moore2007}. This procedure can be generalized to include any space group symmetry in two and three spatial dimensions~\cite{Slager2013} and is expected to become experimentally accessible and relevant in the presence of lattice defects \cite{Ran2009, Teo2010, Juricic2012, Slager2014, Slager2015, Slager2016}. More generally, the interplay of the rich structure of space group symmetries and topology entails an active field of research, providing for new phases and according quasiparticles \cite{Fu2011, BernevigFangGilbert, PhysRevB.93.205104, BernevigHourGlass, Bradlynaaf5037}.

In addition to their role in characterizing topological insulators, lattice symmetries are also vital in describing the phases that emerge on the boundary between topologically distinct phases. A notable example is found in gapless Weyl semimetals \cite{Wan2011}, one of which has been recently  identified experimentally in TaAs \cite{Xu2015, Lv2015a, Lv2015b}. The topological origin of these Weyl phases ensures the presence of specific surface states, called Fermi-Arcs, which connect the band crossings in the bulk. The presence of either three-fold rotations or nonsymmorphic space group symmetries in these materials guarantees that the bulk band crossings cannot be gapped, and it therefore also protects the corresponding Fermi-arcs~\cite{ Young2012, Fang2012, Young2015}.

\begin{table*}[t]\label{Table::classification}
\begin{tabular}{c||lllllllllllllllll}
$\widehat{G}$ & $p1$ & $p2$ & $pm$ & $pg$ & $cm$ & $p2mm$ & $p2mg$ & $p2gg$ & $c2mm$ & $p4$ & $p4mm$ & $p4gm$ & $p3$ & $p3m1$ & $p31m$ & $p6$ & $p6mm$ \\[0.1cm]\hline
$A$ & $\mathbf{Z}^2$ & $\mathbf{Z}^6$ & $\mathbf{Z}^3$ & $\mathbf{Z}$ & $\mathbf{Z}^2$ & $\mathbf{Z}^9$ & $\mathbf{Z}^4$ & $\mathbf{Z}^3$ & $\mathbf{Z}^6$ & $\mathbf{Z}^9$ & $\mathbf{Z}^9$ & $\mathbf{Z}^6$ & $\mathbf{Z}^8$ & $\mathbf{Z}^5$ & $\mathbf{Z}^5$ & $\mathbf{Z}^{10}$ & $\mathbf{Z}^8$
\end{tabular}
\caption{\label{table2d} The complete classification of topological phases within two-dimensional crystals in class $A$ (broken time-reversal, particle-hole, and chiral symmetries). The wallpaper groups $\widehat{G}$ in the first row are denoted in the Hermann-Mauguin notation~\cite{Hahn2002}. The second row denotes the number of integers that need to be specified in order to completely characterise a topological phase in the corresponding wallpaper group.}
\end{table*}

In this article, we expand the use of space group symmetries to provide a simple, but universal, algorithm for identifying and labelling all possible topological phases in class $A$ (i.e. in crystals with broken time-reversal, particle-hole, and chiral symmetries). Essentially, we use elementary representation theory to characterise topologically distinct band structures. In two dimensions this results in the complete list of allowed topological phases shown in Table~\ref{table2d}. The equivalent table in three dimensions can be constructed using the same procedure. Our arguments agree with the mathematical computations in terms of twisted equivariant $K$-theory, as proposed by Freed and Moore~\cite{Freed2013}, thereby elucidating its involved mathematical notions in a straightforward physical setting. 

The algorithm will be worked out in detail below, but can be presented here on a heuristic level. The occupied bands in a crystal are described by Bloch functions on the Brillouin zone (BZ). These functions transform under the crystal symmetry in a particular way which changes as one goes from a generic point in the BZ to a high symmetry point. As there are different ways of reaching such high symmetry points, the transformation rules of Bloch functions need to satisfy gluing conditions which ensure their mutual compatibly~\cite{Wigner1936}. The possible valence band structures in a crystal are thus limited to ones that are consistent with the gluing conditions implied by its crystal symmetry. The way a valence band transforms under crystal symmetries can only be altered by exchanging it with a conduction band, which necessarily involves a closing of the band gap. Since topological phases of matter are defined to be robust to changes that keeps the gap open, an alteration in the transformation properties of the valence band can be interpreted as a topological phase transition. This is analogous to the way changes in more familiar topological invariants, such as the Chern number or TKNN invariant \cite{Thouless1982, Avron1985}, are necessarily accompanied by a closing of the gap. We therefore find that the transformation properties of the valance band characterise its topological phase, and need to be included in the topological classification. We describe the transformation properties of the valence band by a set of integers, which, together with the Chern number, completely specifies the topological phase of any crystal within class $A$.

Because an exchange of valence and conduction bands implies a closing of the gap, we can examine the impact of crystal symmetries on topological phase transitions by studying band inversions at high symmetry points in the first Brillouin zone. Such inversions are accompanied by either a direct transition between topological phases, or the formation of intermediate (Weyl) semimetallic phases. We give a complete analysis of the two-dimensional case, and list all possible types of transitions and intermediate phases consistent with band inversions for the 17 wallpaper groups in Table \ref{transitions}.

The paper is organized as follows. We first  present the example of a specific two-dimensional crystal structure to illustrate the classification scheme on a conceptual level, and to introduce some notation. In Section \ref{sec::genwpg} we then discuss the general case in two dimensions. This will pave the way for the description of possible intermediate phases emerging in topological phase transitions in two dimensions, which is the subject of Section \ref{sec::ps}. We show in Section \ref{sec::gen} that three dimensional topological insulators in class $A$ can be classified using the same scheme after taking into account some additional subtleties. Generalizations to other classes will also be considered in this section. Finally, we discuss our findings and comment on future work. 

\section{Analysis of $p4mm$\label{sec::p4mm}}
\begin{figure}[t]
\centering
\def\svgwidth{4cm}
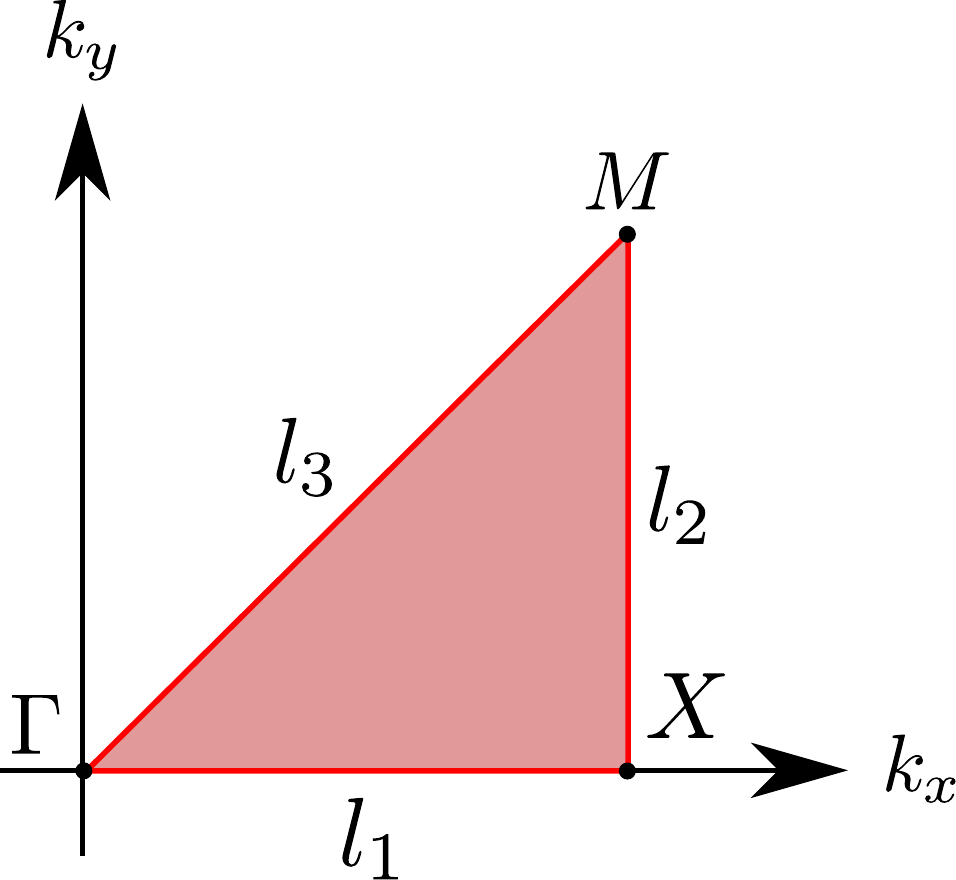
\caption{\label{funddomainp4mm} The fundamental domain (shaded red) $\W$ of the first Brillouin zone. It contains only points that are not related to each other by transformations in the point group $D_4$, as described by equation~\eqref{actionD4}. high symmetry lines are indicated in red, and high symmetry points in black.}
\end{figure}

To give a conceptual description of the proposed classification scheme, we first focus on the particular example of a two-dimensional crystal whose crystal structure falls within the symmorphic wallpaper group $p4mm$. Consider a square array of atoms with lattice spacing $a$, which is set to unity ($a = 1$) in the remainder of this paper. The lattice is spanned by the lattice vectors $\mathbf{t}_1 = (1,0)$ and $\mathbf{t}_2 = (0,1)$. Besides the lattice translations, the crystal is symmetric under all operations that leave a square invariant. These symmetries form the point group $D_4$, which is generated by a reflection $t$ about the $x$-axis and an in-plane $90^{\circ}$-rotation $r$. A general element $g$ of the space group $p4mm$ consists of the combination of a point group element $R$ (centered at the origin) and a translation along a vector $\mathbf{v} = n_1\mathbf{t}_1 + n_2\mathbf{t}_2$. We denote such space group elements as:
\be
g = \{R | n_1 n_2\}.
\ee 
Going to momentum space, we again find a square lattice in terms of the two reciprocal lattice vectors $\mathbf{g}_1 = 2\pi(1,0)$ and $\mathbf{g}_2 = 2\pi(0,1)$. The first Brillouin zone is a square with $-\pi \leq k_{x,y} \leq \pi$. Opposing edges of the square are equivalent and may be identified, giving the first Brillouin zone the topology of a torus. The presence of a crystal symmetry allows us to consider only part of the first Brillouin zone. This is because a space group element $g = \{R | \mathbf{v}\}$ transforms an electronic Bloch function at momentum $\mathbf{k}$ to one with momentum $R\cdot \mathbf{k}$. The band structure is therefore already fully specified if it is determined just for those points that are not related to each other by the action of the point group. This region of wave vectors is known as the fundamental domain, $\W$. The $D_4$ symmetry operations $r$ and $t$ in our present example affect the momenta of states within the first Brillouin zone according to:
\be\label{actionD4}
r\cdot (k_x,k_y) = (-k_y,k_x),\quad t \cdot (k_x,k_y) = (k_x,-k_y).
\ee
The fundamental domain thus consists of the region with momentum values $0\leq k_x \leq \pi,\;0\leq k_y \leq k_x$, as shown in Figure \ref{funddomainp4mm}.

Within the fundamental domain, there are special wave vectors that are mapped onto themselves by some or all of the operations that make up the point group $D_4$. The electronic states with momenta corresponding to such special wave vectors are then eigenstates of the subgroup of operations in $D_4$ which leave their momentum invariant. These special wave vectors may appear at high symmetry points and lines in the first Brillouin zone, which are shown in Figure \ref{funddomainp4mm} and listed in Table \ref{fixedpntsp4mm}. 

As an example, consider the origin $\G = (0,0)$. The momentum of Bloch states at this point in the first Brillouin zone is held fixed under both $r$, and $t$, and any combination of reflections and rotations. The same is true for $M = (\pi,\pi)$, because under $r$ and $t$ this point is mapped onto itself modulo a reciprocal lattice vector. The presence of reflections in the group $D_4$, also allows entire lines in the first Brillouin zone to be left invariant under some of the point group operations. One readily verifies that $l_1 = (k_x,0)$ is left invariant by $t$, while $l_2 = (\pi,k_y)$ and $l_3 = (k_x,k_x)$ are invariant under the action of $r^2t$ and $rt$, respectively. At the intersection of $l_1$ and $l_2$ we find the point $X = (\pi,0)$, which must be left invariant under both the symmetries that leave $l_1$ unaffected, and the symmetries that keep $l_2$ fixed.

\begin{table}
\begin{tabular}{l||ll}
 & \quad $\mathbf{k}$ & little co-group  $G^{\mathbf{k}}$ \\\hline
$\G$ & $(0,0)$ & $D_4$\\\hline
$M$ & $(\pi,\pi)$ & $D_4$\\\hline
$X$ & $(\pi,0)$ & $\mathbf{Z}_2 \times \mathbf{Z}_2 = \{1,r^2, t, r^2t\}$\\\hline
$l_1$ & $(k_x,0)$ & $\mathbf{Z}_2 = \{1,t\}$\\\hline
$l_2$ & $(\pi,k_x)$ & $\mathbf{Z}_2 = \{1,r^2t\}$\\\hline
$l_3$ & $(k_x,k_x)$ & $\mathbf{Z}_2 = \{1,rt\}$\\\hline
$\text{int}(\W)\;$ & $(k_x,k_y)$ & $\{1\}$\\\hline
\end{tabular}
\caption{\label{fixedpntsp4mm} The little co-groups contained within the space group $p4mm$. These little co-groups consist of all symmetry operations which keep the momentum of a particular high symmetry point or line fixed. The fundamental domain $\W$ in this case contains momentum points with $0<k_x<\pi$ and $0<k_y<k_x$, and its interior is denoted $\text{int}(\W)$.}
\end{table}

The symmetry transformations, or elements of the point group, which leave a particular high symmetry point $\mathbf{k}$ invariant, by themselves also form a group. This group is referred to as the \emph{little co-group} $G^{\mathbf{k}}$. From the arguments above, we see that the little co-groups at the high symmetry points $\G$ and $M$ are equal to the full point group, $G^{\G} = G^{M}= D_4$, because all symmetry transformations in $D_4$ leave these points invariant. Similarly, the high symmetry lines are left invariant only by a single transformation, and their little co-groups are thus $G^{l_i} = \mathbf{Z}_2$, with $i=1,2,3$. Notice that these $G^{l_i}$ are generated by different transformations along different high symmetry lines, but that the resulting group structure is the same in each case. The point $X$ finally, combines the symmetries of two lines, and has little co-group $G^{X} = \mathbf{Z}_2\times\mathbf{Z}_2$. All little co-groups, and associated symmetry transformations, are listed in Table~\ref{fixedpntsp4mm}.

The little co-groups at high symmetry points may necessitate bands at those points to become degenerate. This can be seen directly from the way a space group element acts on the Bloch function $\psi_{\mathbf{k},i}(\mathbf{r})$ with band index $i$, position $\mathbf{r}$, and wave vector $\mathbf{k}$:
\be\label{actionbloch}
\{R|\mathbf{v}\}\cdot \psi_{\mathbf{k},i}(\mathbf{r}) = \sum_{j} D^{\mathbf{k}}_{ij}(\{R|\mathbf{v}\})\psi_{R\mathbf{k},j}(\mathbf{r}-\mathbf{v}).
\ee
The point group element $R$ transforms a Bloch function with momentum $\mathbf{k}$ to a Bloch function with momentum $R\cdot\mathbf{k}$.  Bloch functions with different band indices $j$, but equal momentum $\mathbf{k}$, may be mixed by the matrix $D_{ij}^{\mathbf{k}}(\{R | \mathbf{v}\})$. This matrix is a representation of the space group element $\{ R | \mathbf{v}\}$, which for generic points in the fundamental domain away from high symmetry locations, is usually just a phase factor (i.e. a one-dimensional representation). The elements of the space group for which $\mathbf{v}$ is a pure lattice translation, combine Bloch functions $\psi_{R\mathbf{k},j}(\mathbf{r}-\mathbf{v})$ that differ from the functions $\psi_{R\mathbf{k},j}(\mathbf{r})$ by pure phase factors. Space groups consisting only of these types of elements, such as $p4mm$, are called symmorphic, and in dealing with these space groups we only need to consider ordinary representations of the point groups when determining the $D_{ij}^{\mathbf{k}}(\{R | \mathbf{v}\})$. For nonsymmorphic space groups, which contain screw axes or glide planes, and have elements with $\mathbf{v}$ a fraction of a lattice translation, projective representations of the point groups need to be taken into account. These will be considered in detail in Appendix \ref{sec::NSyms}. 

Since the Hamiltonian is symmetric with respect to the space group of the lattice, eigenstates of space group elements must also be eigenstates of energy. Equation~\eqref{actionbloch} shows the eigenstates of $\{ R | \mathbf{v}\}$ to be linear combinations of $\psi_{R\mathbf{k},j}(\mathbf{r})$, so these states must all have equal energy. For states at high symmetry points, such that $R\cdot \mathbf{k} = \mathbf{k}$, all bands at $\mathbf{k}$ connected by  $D_{ij}^{\mathbf{k}}(\{R | \mathbf{v}\})$ are then necessarily degenerate. This conclusion can also be expressed on the level of the Hamiltonian itself. If the full Hamiltonian is written as a sum of Bloch Hamiltonians $H(\mathbf{k})$, the action of the crystal symmetries can be described by:
\bea\label{actionHam}
\rho(R) H(\mathbf{k}) \rho(R)^{-1} &=& H(R\cdot\mathbf{k}) \notag \\
\left[\rho(R), H(\mathbf{k}) \right] &=& 0 ~~~\text{if}~R\cdot \mathbf{k} = \mathbf{k}.
\eea
Here $\rho(R)$ is a (matrix) representation of the point group element $R$, or equivalently, an operator enacting its symmetry transformation. The Bloch Hamiltonian commutes with the elements of the little co-group at high symmetry points and lines. At these locations, the eigenfunctions of the Bloch Hamiltonian are thus also eigenfunctions of the elements of the little co-group. Conversely, the collection of states in the valence band with momentum $\mathbf{k}$, forms a representation of the little co-group $G^{\mathbf{k}}$. This representation consists of irreducible representations of $G^{\mathbf{k}}$, which represent either individual bands (one-dimensional irreducible representations) or sets of necessarily degenerate bands (higher-dimensional irreducible representations). Determining the irreducible representations of $G^{\mathbf{k}}$ can thus be interpreted as a recipe for constructing the entire set of valence bands at high symmetry locations. As we will see in the following, however, it is necessary to impose additional constraints when considering the structure of the valence bands throughout the first Brillouin zone.

These constraints come from the fact that representations along high symmetry lines need to connect properly, i.e continuously,  to representations at their endpoints, the high symmetry points. That is, if a Bloch state has a certain eigenvalue for a symmetry transformation on a high symmetry line, that eigenvalue cannot suddenly change at the endpoint of the line. For the case of $p4mm$, the bands at $\G$ and $M$ form a representation of $D_4$, and at $X$ of $\mathbf{Z}_2\times\mathbf{Z}_2$. Along the lines $l_i$ connecting these three points, the bands need to form a representation of $\mathbf{Z}_2$. Symmetry transformations making up a $\mathbf{Z}_2$ group structure always have eigenvalues $\pm 1$, so that the eigenstates along $l_i$ can be either even $(+)$ or odd $(-)$ under the transformation:
\bea
R_i \cdot \ket{u_{\mathbf{k}}, \pm} &=& l_{i,\pm}(R_i) \ket{u_{\mathbf{k}}, \pm} 
\eea
Here $R_i$ is an element of the little co-group $\mathbf{Z}_2$ along $l_i$, and $\ket{u_{\mathbf{k}},  j}$ represents a state at momentum $\mathbf{k}$ with reflection eigenvalue $\pm 1$. The eigenvalues $l_{i,\pm}(R_i)=\pm 1$ in general are representations of the little co-group $\mathbf{Z}_2$ along $l_i$. Since the representations are one-dimensional (i.e. they apply to a non-degenerate band), they can be replaced by their characters, which equal  the eigenvalues $\pm 1$. In the general case of higher dimensional representations, or having a degenerate set of Bloch functions, the representations become matrices.
\begin{table}[t]
\begin{center}
\begin{tabular}{l||cc}
& $\{1\}$ & $\{t\}$ \\\hline
$l_{i,+}$ & $1$ & $1$ \\\hline 
$l_{i,-}$ & $1$ & $-1$ \\\hline 
\end{tabular}
\end{center}
\caption{\label{charactertableZ2} The character table of $\mathbf{Z}_2$. The irreducible representations along the line $l_i$ are denoted by $l_{i,\pm}$. Here $\pm$ signifies whether states are respectively even or odd under the symmetry transformation.}
\end{table}
\begin{table}
\begin{center}
\begin{tabular}{l||ccccc}
& $\{1\}$ & $\{r^2\}$ & $\{r,r^3\}$ & $\{t,r^2t\}$ & $\{rt,r^3t\}$\\\hline
$\G_0$ & $1$ & $1$ & $1$ & $1$ & $1$\\\hline 
$\G_1$ & $1$ & $1$ & $1$ & $-1$ & $-1$\\\hline 
$\G_2$ & $1$ & $1$ & $-1$ & $1$ & $-1$\\\hline 
$\G_3$ & $1$ & $1$ & $-1$ & $-1$ & $1$\\\hline 
$\G_4$ & $2$ & $-2$ & $0$ & $0$ & $0$\\\hline 
\end{tabular}
\end{center}
\caption{\label{charactertableD4}The character table of $D_4$ at $\G$. The irreducible representations are denoted by $\G_i$. The little co-group at $M$ is also $D_4$ and has the same character table, but the irreducible representations are denoted by $M_i$.}
\end{table}
\begin{table}
\begin{center}
\begin{tabular}{l||cccc}
& $\{1\}$ & $\{r^2\}$ & $\{t\}$ & $\{r^2t\}$ \\\hline
$X_0$ & $1$ & $1$ & $1$ & $1$ \\\hline 
$X_1$ & $1$ & $1$ & $-1$ & $-1$ \\\hline 
$X_2$ & $1$ & $-1$ & $1$ & $-1$ \\\hline 
$X_3$ & $1$ & $-1$ & $-1$ & $1$ \\\hline 
\end{tabular}
\end{center}
\caption{\label{charactertableZ2Z2} The character table of $\mathbf{Z}_2\times\mathbf{Z}_2$ at $X$. Irreducible representations in this case are denoted by $X_i$.}
\end{table}

If we now follow a particular band along a high symmetry line $l_i$ towards its endpoint, the eigenvalues of the symmetry transformation are preserved along the line. On the high symmetry point at the end of the line, the state is symmetric under more symmetries, and gains some additional quantum numbers describing those, but it retains the eigenvalue that it carried along the line. We are thus restricted in the choice of representation on the high symmetry points by the representations along the high symmetry lines. In other words, when we follow any band towards an endpoint of $l_i$, its representation is lifted to either a representation of $D_4$ or of $\mathbf{Z}_2\times\mathbf{Z}_2$, depending on the endpoint. In terms of character tables, this means that the characters in the character table corresponding to the common elements at $l_i$ and the high symmetry points need to agree. For example, suppose a band transforms as $l_{1,-}$, i.e. it is odd under $t$ along $l_1$. As the entire line $l_1$ is held fixed by $t$, the action of $t$ at the endpoint $X$ must be the same as its action along $l_1$. Consulting the character table~\ref{charactertableZ2Z2},  we see that at $X$, the band must thus transform as either $X_1$ or $X_3$. 

At $\G$, the other endpoint of $l_1$, the band should similarly remain odd under $t$. According to the character table~\ref{charactertableD4}, we then see that at $\G$, the band must transform as either $\G_1$ or $\G_3$. In fact, the two-dimensional representation $\G_4$ is also a possibility, as long as there is an additional even band along $l_1$. In that case, the even and odd bands becoming degenerate at $\G$ would be consistent with $t$ being represented in $\G_4$ by a two-dimensional matrix with eigenvalues $1$ and $-1$ (the character for $\G_4$ in table~\ref{charactertableD4} is the sum of eigenvalues). At $\G$ the two band then form a doublet of $D_4$. 

Repeating this analysis for even bands along $l_1$, the entries in tables \ref{charactertableZ2Z2} and \ref{charactertableD4} for the conjugacy classes $\{t\}$ and $\{t,r^2t\}$ respectively, need to be $1$. Hence, even bands along $l_1$ end in $X_0$ or $X_2$ at $X$ and go to $\G_0$, $\G_2$ or $\G_4$ at $\G$. Applying similar constraints to all high symmetry locations ensures a consistent representation of the entire set of valence bands throughout the fundamental domain. 

Completing the list of which representations along lines $l_i$ enhance to which representations at the endpoints $\G$, $X$, and $M$ results in Table~\ref{enhance}. It shows for example that starting from a given representation at $\G$, only certain representations at the other high symmetry locations are allowed. Below we will use this information to define a set of integers that specifies the representation of the complete set of valance bands. These integers then characterize the topological phase in space group $p4mm$, modulo Chern numbers.
\begin{table}[t]
\begin{center}
\begin{tabular}{l||ll}
 & group enhancement & rep. enhancement\\\hline 
$l_1$ & $D_4 \leftarrow \mathbf{Z}_2$ & $\G_0, \G_2, \G_4 \leftarrow l_{1,+}$ \\\hline
& $D_4 \leftarrow \mathbf{Z}_2$ & $\G_1, \G_3, \G_4 \leftarrow l_{1,-}$ \\\hline
& $\mathbf{Z}_2 \times \mathbf{Z}_2 \leftarrow \mathbf{Z}_2$ & $X_0, X_2 \leftarrow l_{1,+}$\\\hline
& $\mathbf{Z}_2 \times \mathbf{Z}_2 \leftarrow \mathbf{Z}_2$ & $X_1, X_3 \leftarrow l_{1,-}$\\\hline
$l_2$ & $D_4 \leftarrow \mathbf{Z}_2$ & $M_0, M_2, M_4 \leftarrow l_{2,+}$ \\\hline
& $D_4 \leftarrow \mathbf{Z}_2$ & $M_1, M_3, M_4 \leftarrow l_{2,-}$\\\hline
& $\mathbf{Z}_2 \times \mathbf{Z}_2 \leftarrow \mathbf{Z}_2$ & $X_0, X_3 \leftarrow l_{2,+}$\\\hline
& $\mathbf{Z}_2 \times \mathbf{Z}_2 \leftarrow \mathbf{Z}_2$ & $X_1, X_2 \leftarrow l_{2,-}$\\\hline
$l_3$ & $D_4 \leftarrow \mathbf{Z}_2$ & $\G_0, \G_3, \G_4 \leftarrow l_{3,+}$ \\\hline
& $D_4 \leftarrow \mathbf{Z}_2$ & $\G_1, \G_2, \G_4 \leftarrow l_{3,-}$\\\hline
& $D_4 \leftarrow \mathbf{Z}_2$ & $M_0, M_3, M_4 \leftarrow l_{3,+}$ \\\hline
& $D_4 \leftarrow \mathbf{Z}_2$ & $M_1, M_2, M_4 \leftarrow l_{3,-}$\\\hline
\end{tabular} 
\end{center}
\caption{\label{enhance} The list of consistency relations between representation along high symmetry lines $l_i$ and possible representations at their endpoints $\G$, $X$, and $M$.}
\end{table}

\subsection{Counting the topological phases protected by $p4mm$\label{p4mmcounting}}
The topological phases we would like to characterize, are defined to be phases of matter which are stable under any deformations that do not close the gap between valence and conductions bands, and that do not change the crystal symmetry. Deformations that do close the gap, necessarily cause either the representation of the set of valance bands or the Chern numbers to change. This means that a topological phase can be uniquely specified by the representation of its set of valance bands and its Chern numbers. For the specific case of the space group $p4mm$, there are no Chern-numbers \cite{BernevigFangGilbert} due to the reflection symmetry in $D_4$, and so its topological phases within class $A$ are completely specified once the representation of the set of valance bands is known. In table \ref{enhance} we already identified constraints on the allowed representations, which we will now employ to classify the possible topological phases of $p4mm$.

The representation of the set of valance bands, denoted by $\mathcal{V}$, is built out of a number of irreducible representations at each high symmetry point. To specify $\mathcal{V}$, we therefore simply count the number of bands in each irreducible representation at the high symmetry points, subject to the constraints in Table \ref{enhance}. As can be seen from the character tables \ref{charactertableD4}, \ref{charactertableZ2Z2} there are five irreducible representations at both $\G$ and $M$, and four at $X$. This results in $14$ integers, $n_i^{\mathbf{k}}$, signifying how many bands there are at $\mathbf{k}$ transforming under the representation labeled $i$ at that point. For example, the integer $n_2^X$ indicates the number of bands at $X$ transforming as $X_2$.

We can then consult Table \ref{enhance} to see that only certain representations at $\G$, $X$ and $M$ are possible depending on whether the bands are odd or even along the connecting lines $l_i$. This relates the integers $n_i^{\mathbf{k}}$ at different high symmetry points to each other. For instance, the number of even bands along $l_1$, $n^0_{l_1}$ must be equal to the combined number of bands in $\G_0$, $\G_1$ and $\G_4$ at $\G$,
\be\label{rel1}
n_{l_1}^0 = n_0^{\G} + n_2^{\G} + n_4^{\G}.
\ee
Moreover, going to the other endpoint, $X$, the number of even bands must equal the sum of those in $X_0$ and $X_2$. The combination of the two relations between the number of even representations along the high symmetry lines and the combined numbers of representations at its endpoints then implies a direct relation between the high symmetry points:
\be\label{firstrel}
n_0^{\G} + n_2^{\G} + n_4^{\G} = n_0^X + n_2^X.
\ee
Repeating these steps for the odd bands along $l_1$, we find a similar relation,
\be
n_1^{\G} + n_3^{\G} + n_4^{\G} = n_1^X + n_3^X.
\ee
The integer $n_4^{\G}$ specifying the number of bands in the two-dimensional representation $\G_4$ appears in both sets of relations, because a doublet at $\G$ must split into both an even and odd band along $l_1$. The analysis for the remaining high symmetry lines $l_2$ and $l_3$ is similar and yields the relations:
\bea
n_0^{\G} + n_3^{\G} + n_4^{\G} &=& n_0^M + n_3^M + n_4^M\\
n_1^{\G} + n_2^{\G} + n_4^{\G} &=& n_1^M + n_2^M + n_4^M\\
n_0^{M} + n_2^{M} + n_4^{M} &=& n_0^X + n_3^X\\
n_1^{M} + n_3^{M} + n_4^{M} &=& n_1^X + n_2^X.\label{lastrel}
\eea

The six relations between integers $n_i^{\mathbf{k}}$ show that they cannot be chosen independently, and they thus reduce the number of integers required to specify the complete representation of the set of valance bands. In fact, only five of the six relations are independent from each other. That is, the rank of the system of equations relating different integers $n_i^{\mathbf{k}}$ has rank $m = 5$. This implies that $14-m=9$ integers need to be specified to characterize the set of valance bands. These nine integers completely fix how many valence bands there are and under which representations they transform on all high symmetry points in the fundamental domain. We thus conclude that the topological phases in class $A$ protected by $p4mm$ space group symmetry can be classified by a set of nine integers, i.e. by elements of $\mathbf{Z}^9$.

\section{General wallpaper group \label{sec::genwpg}}
The method exemplified by our analysis of the wallpaper group $p4mm$ can be applied in the same way to all wallpaper groups. The first step is always to determine the fundamental domain $\W$. After that, the point group operations are used to identify high symmetry points and lines as well as their corresponding little co-groups. The correspondence between characters of the little co-groups along high symmetry lines, and those at the high symmetry endpoints, then yield the allowed combinations of their irreducible representations, akin to the example of Table \ref{enhance}. To specify a representation of the full set of valance bands, an integer $n_i^{\mathbf{k}}$ should be assigned to each irreducible representation $i$ at high symmetry point $\mathbf{k}$, which specifies the number of valence bands for those values of $i$ and $\mathbf{k}$. The previously listed relations between irreducible representations along high symmetry lines and their endpoints, can then be converted into a set of $m$ independent relations between the integers $n_i^{\mathbf{k}}$. A representation of the complete set of valance bands is specified by $n-m$ integers, where $n$ is the total number of integers $n_i^{\mathbf{k}}$ one started with. Finally, we need to consider Chern numbers. These topological invariants can only be present in wallpaper groups that do not contain reflections, because the Berry curvature is odd under reflection. For groups that do allow a Chern number, this one additional integer should be added to the set of $n_i^{\mathbf{k}}$ in order to have a complete specification of the set of valence bands.

Four of the $17$ wallpaper groups, $pg$, $p2gg$, $p2mg$ and $p4gm$, i.e. the nonsymmorphic ones, need special attention. In these cases, the representations of the little co-group become projective representations. This is a consequence of the fact that we cannot separate their point group actions $R$ from the translations $\mathbf{v}$, and as a result an additional phase factor needs to be accounted for in the analysis \cite{bradleycracknell, Young2012, Young2015, BernevigHourGlass}. As long as this subtlety is properly taken into consideration however, the procedure outlined above can still be applied in precisely the same way. This again enables us to identify the set of integers needed to completely specify the topological phase given the space group symmetry. A detailed discussion of the procedure for nonsymmorphic groups is given in Appendix \ref{sec::NSyms}. 

Table \ref{table2d} collects our results, and classifies all topological phases in class $A$ for any of the $17$ wallpaper groups. It exactly agrees with the mathematical computation in terms of $K$-theory as proposed by Freed and Moore in \cite{Freed2013}. This mathematical theory is the formal framework for classifying topological phases of gapped free fermions, and should therefore match our results. The agreement between  $K$-theory calculations and table~\ref{table2d} based on our combinatorial arguments, can be made explicit using the results of \cite{luck2000, Yang1997, Phillips1989, RoyerFreed}. As the comparison involves some intricate mathematical details, however, we refer the interested reader to Appendix \ref{sec::Mat}.

\section{Intermediate phases\label{sec::ps}}
Having found a classification of all possible topological phases in two dimensional crystals, we can now also analyse allowed transitions between them. The symmetry-protected phases classified by our band structure combinatorics scheme are stable as long perturbations do not close the band gap and do not break the lattice symmetry. Lattice distortions, such as defects or impurities, are examples of perturbations that locally break the space group symmetry. A small number of isolated defects or impurities, however, will not significantly affect the properties of the topological phase, since they can be expected to only change the electronic band structure in a negligible way. The topological transitions that may result from breaking or altering the lattice symmetry on a more global level will be left for future work.

In this section we will focus instead on topological transitions that do not alter the space group symmetry of the crystal lattice. Such transitions are driven either by changes in the band filling or by band inversions. Changing the band filling by adding electrons or holes to a material, can turn any insulator into a metal and therefore has no direct topological significance. We therefore focus exclusively on transitions driven by band inversions at fixed filling from here on. Such inversions can occur anywhere in the first Brillouin zone, and can turn the insulator into a semimetal, hosting degeneracies at arbitrary points in the first Brillouin zone.  For generic values of the momentum value $\mathbf{k}$ at which an inversion occurs however, the resulting degeneracies in two-dimensional materials will generically be lifted and an avoided crossing is realised instead. This can be seen directly from the counting arguments in our combinatoric procedure. Consider the two bands that are being inverted. They can be parameterised by three Pauli matrices, which means we need to tune three coefficients to obtain a degeneracy. In two dimension there are only two components of the momentum that can be tuned, and states at generic value of $\mathbf{k}$ are not eigenstates of any symmetries that could be used to force the remaining third parameter to take a specific value. The implications is that nodes do not occur at generic momentum values without fine-tuning. For band inversions on high symmetry lines on the other hand, a point group operation keeping that line fixed, such as the mirror operation for lines in $p4mm$, will cause the momenta of the two nodes resulting from the inversion to be constraint to the high symmetry line. These nodes are still unstable however, because they can be moved along the mirror line by a perturbation of the system in such a way that pairs of nodes annihilate one another. By doing this, the band inversion can be undone, so no topological phase transition could have occurred as a result of the inversion on the high symmetry line. The only interesting case in two dimensions then, involves band inversions at high symmetry points. These types of inversions result in single nodes along high symmetry lines in the fundamental domain. They are much more stable than the other cases, because there are no pairs of nodes on the high symmetry lines which can mutually annihilate. To get rid of single nodes, they must be moved all the way to the end of the high symmetry line away from where the band inversion was performed. There, the node can then meet up with a symmetry related  partner which is simultaneously moved along a related high symmetry line outside the fundamental domain, and be annihilated. Because of the relative stability of single nodes, semimetals in class $A$ resulting from a space group symmetry-preserving phase transition will generically be due to inversions at high symmetry points. 

One appealing aspect of the topological classification in terms of band structure combinatoric, presented here, is that the topological indices $n_i^{\mathbf{k}}$ precisely reflect the special role of band inversions at high symmetry points. After all, these are the only operations that can alter the indices and thereby cause a topological phase transition while preserving the space group symmetry. 

To study band inversions in more detail we again begin by focussing on the example of space group $p4mm$. In this case, three types of band inversion can be distinguished, depending on whether single or doubly degenerate bands are inverted. In terms of representation theory, an inversion at a high symmetry point corresponds to changing the representation of the set of valance bands, given by the set of indices $n_i^{\mathbf{k}}$. At the high symmetry point the set of valance bands consists of bands that are either non-degenerate, or stick together in doubly degenerate pairs. These correspond to one- and two-dimensional irreducible representations of the little co-group respectively. The effect of an inversion amounts to interchanging irreducible representations within the set of valance bands with those from the set of conduction bands. The key ingredient in understanding the result of these inversions, is to determine whether the band crossings created along $l_i$ may be avoided or not. This can be done by determining whether or not the eigenvalues corresponding to the symmetry transformations along $l_i$ are  the same for the bands involved in the crossing. If two bands with equal eigenvalues cross each other, a perturbation can always be introduced which causes a splitting of the bands without affecting the symmetry of the system, and the crossing will in practice thus be avoided. If the eigenvalues of the bands differ from one another however, splitting them could only be done by operations that do not respect the space group symmetry, and such nodes are thus symmetry-protected, and stable.

\subsection{Nodes along high symmetry lines}
\begin{figure}[t]
\centering
\def\svgwidth{8cm}
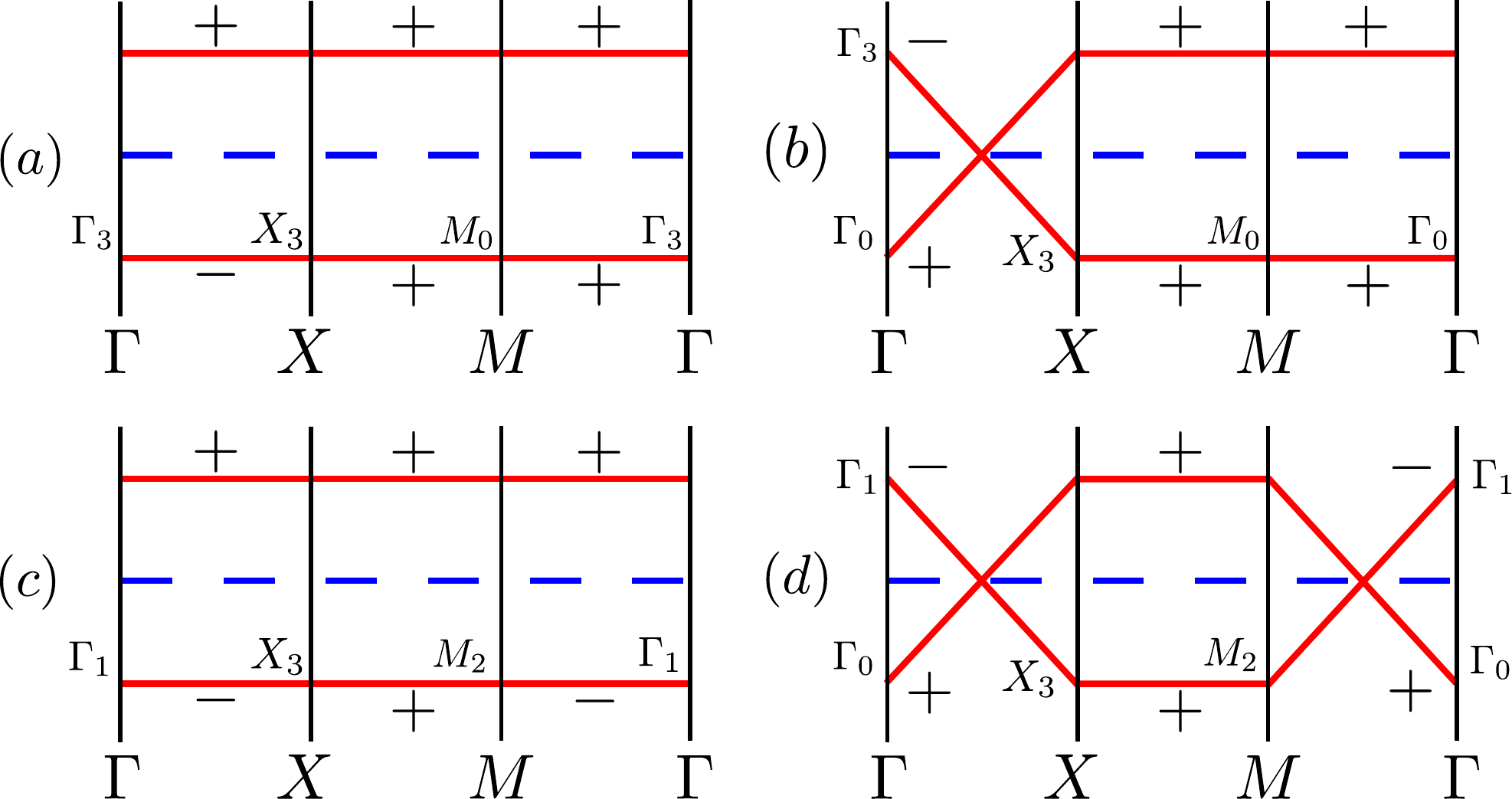
\caption{\label{Weyl} Possible effects of a  band inversion at $\G$ on band structure topologies with $p4mm$ symmetry. Red lines are electronic bands, which are flattened for clarity. The dashed blue line indicates the Fermi energy. (a) A topologically insulating state with one valance band and one conduction band. The representations, or $\pm1$ eigenvalues, along high symmetry lines are indicated. Representations at high symmetry points are indicated for the valance band only, as the conduction band is in the trivial representation at all high symmetry points. This configuration is gapped everywhere, and has $N_{\pm}^{l_1} = 0$. (b) Upon an inversion at $\G$, two band will cross along $l_1$, creating a Weyl cone. The crossing is characterised by $N_{\pm}^{l_1} = \pm 1$. Note that along $l_3$, a crossing has been avoided. (c) A topologically insulating state with a valance band that is odd on both $l_1$ and $l_3$. (d) An inversion at $\G$ in this case results in Weyl nodes along both $l_1$ and $l_3$, characterised by $N_{\pm}^{l_1} = \pm 1 = - N_{\pm}^{l_3}$.}
\end{figure}

Consider a system with $p4mm$ space group symmetry, which contains two non-degenerate bands, one above and one below the Fermi level, as shown in Figure~\ref{Weyl}(a). Let us also assume that the conduction band is in the trivial representation at $\G$, $M$ and $X$, meaning that it is even along all $l_i$. In contrast, assume that the valence band is in a non-trivial representation, i.e an odd one, along one of the $l_i$, say $l_1$. An inversion at $\G$ or $X$ will now create a symmetry-protected Weyl node along $l_1$, because the band crossing that it induces cannot be avoided while respecting the lattice symmetry. The nodes created along other high symmetry lines contain bands with equal eigenvalues for the symmetry transformation, and can thus be trivially avoided or split. The stable state after the inversion is thus one in which only the protected node remains, as shown in Figure \ref{Weyl}(b). Reflection symmetry fixes the location of this node to lie on $l_1$, but it can be moved along the high symmetry line in agreement with our previous counting of the number of tunable parameters. 

A second scenario we may consider, is that the valance band originally is odd along two high symmetry lines, while keeping the conduction band even everywhere. In that case, two symmetry-protected nodes will be produced upon performing an inversion. Examples of this can be seen in Figures \ref{Weyl}(c) and (d). In the case of the $p4mm$ space group, there are no more possibilities to consider, because there are no points in the fundamental domain at which more than two high symmetry lines meet.

In both cases considered, the inversion of two non-degenerate bands gives rise to an intermediate semimetallic phase. This phase can be turned into a different topological insulating phase from the original one, by performing a second inversion at the other end of the high symmetry line(s) containing Weyl nodes. In the situation shown in Figures \ref{Weyl}(b) and (d), this means inverting the bands at $X$ and at $X$ and $M$, respectively.  Performing the second inversion is equivalent to tuning the location of the node to the other end of the high symmetry line, where it then annihilates with a symmetry-related partner from outside the fundamental domain.

In the intermediate semimetallic phase, the dispersion near the nodes is generically linear. Intuitively, this may be expected from the fact that the two bands crossing each other cannot be connected by perturbations that respect the symmetry. Formally, it can be seen from the fact that at the crossing point, the Hamiltonian forms a tensor product representation of the two intersecting bands. Combining this with the fact that the Hamiltonian can always be expanded in the form $H=a+b_ik_i+c_{ij}k_ik_j+\dots$, where the linear term represents a vector representation of the corresponding group, it is clear that if the direct sum decomposition of the tensor product for the crossing bands contains a vector representation, a linear dispersion will generically be present. This is the case for a crossing of two bands with different eigenvalues of a $\mathbf{Z}_2$ symmetry.

The intermediate semimetallic phase can be further characterised using the analysis of Section \ref{sec::p4mm}. An inversion changes the number of valance bands in a particular representation, and in doing so, introduces connections between a valence band at one high symmetry point, and a conduction band at another. As a result, some of the relations \eqref{firstrel} through \eqref{lastrel} will be violated. Equivalently, we can also say that whenever these relations are not satisfied there {\it must} be a crossing of valence and conduction bands somewhere in $\W$. since the electronic filling is assumed to be conserved, the degeneracy always happens at the Fermi energy.  

The violations of the relations between numbers of representations can be quantified. Along $l_1$, for example, they can be specified by the integers $N_{\pm}^{l_i}$, with:
\bea\label{Weylcharac1}
N^{l_1}_+ &=& n_0^{\G} + n_2^{\G} + n_4^{\G} - n_0^X - n_2^X,\\\label{Weylcharac2}
N^{l_1}_- &=& n_1^{\G} + n_3^{\G} + n_4^{\G} - n_1^X - n_3^X.
\eea
For the insulating phase in Figure \ref{Weyl}(a), it is clear that $N_{\pm}^{l_1} = 0$, signifying this is a topological insulator. The band structure after the inversion however, shown in Figure \ref{Weyl}(b), has $N^{l_1}_{\pm} = \pm 1$ and $N^{l_1}_{\pm} = \mp 1$, and is a Weyl semimetal. In general, we can be certain a node is present whenever
\be\label{Weylcharac}
N^{l_i}_{+} = - N^{l_i}_{-},
\ee
because in order to get Weyl nodes, a given number of even bands always has to be interchanged with the same number of odd bands. 

For $p4mm$, there are six distinct semimetallic phases, three with a node along one of the high symmetry lines, and another three with a node along two of the high symmetry lines. From equations \eqref{Weylcharac1} and \eqref{Weylcharac2} it is clear that two consecutive inversions at adjacent high symmetry points result in a topologically insulating state with $N_{\pm} = 0$, as anticipated. 

\subsection{Nodes at high symmetry points}
\begin{figure}[t]
\centering
\def\svgwidth{6cm}
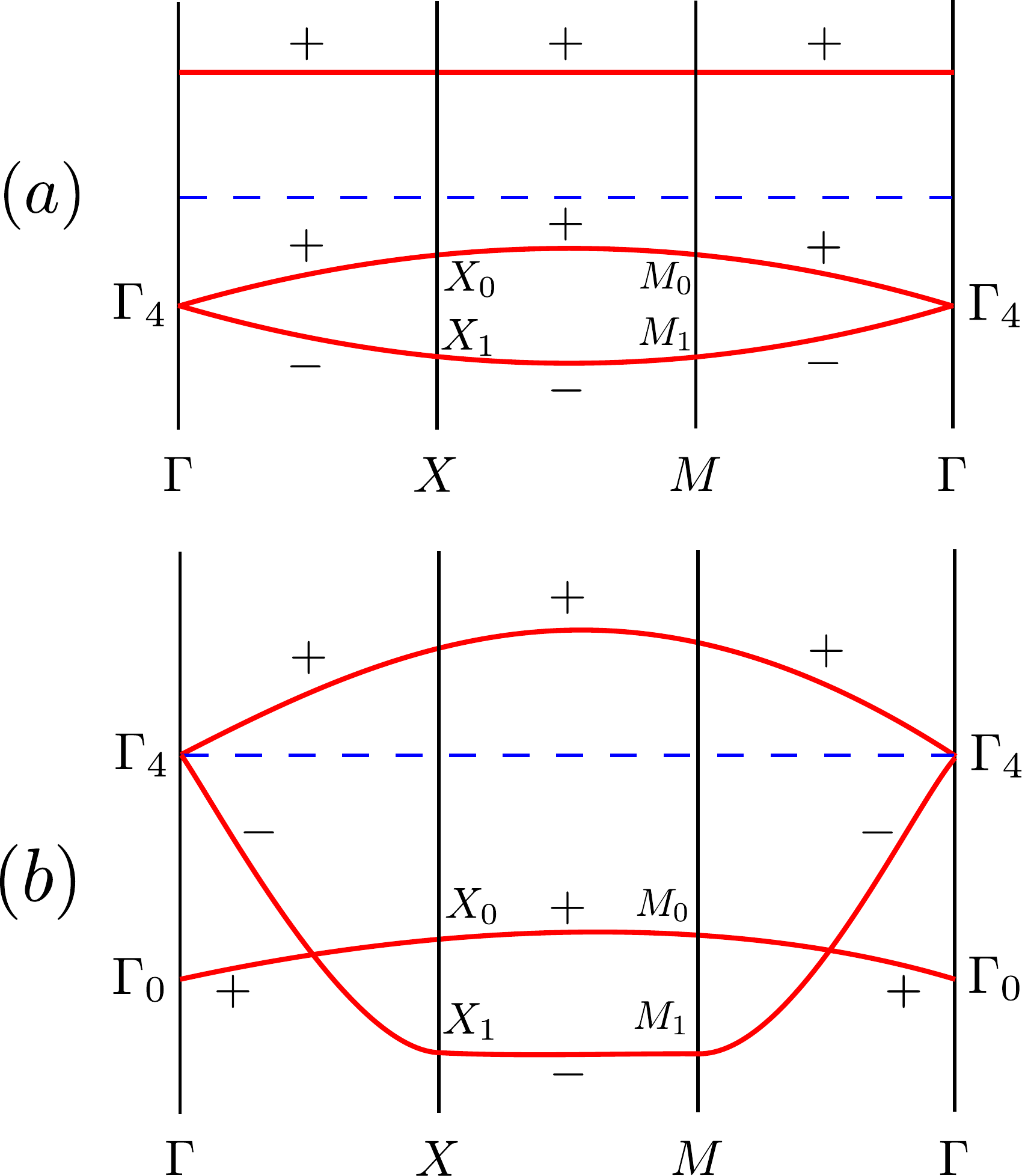
\caption{\label{WeylAtPnt} Inverting a non-degenerate band with a doubly degenerate one, at fixed filling. (a) A topologically insulating state with two valance bands and one conduction band. The valance bands are degenerate at $\G$, and $N_{\pm}^{l_i} = 0$ for all $l_i$. (b) A band inversion at $\G$ at fixed filling requires the number of filled bands after the inversion to still be two everywhere. This implies that at $\G$, a degeneracy must appear at the Fermi level. This intermediate phase is characterized by $N_+^{l_1} = - N_+^{l_3} = 1$, $N_-^{l_1} = N_{-}^{l_3}= 0$.}
\end{figure}

Besides having nodes along high symmetry lines, it is also possible for Weyl nodes to be fixed at high symmetry points, akin for example to the nodes in graphene \cite{Neto2009}. Within our analysis of the possible phases respecting $p4mm$ space group symmetry, these types of nodes result from interchanging a non-degenerate with a doubly degenerate band. Precisely such an inversion occurs for example in the well-known HgTe systems \cite{Bernevig2006, Koning2007}. As an example, consider again a non-degenerate conduction band that is even along all high symmetry lines, and a set of two valance band that are degenerate at $\G$, i.e. the valance band is in the representation $\G_4$ at $\G$, as shown in Figure \ref{WeylAtPnt}(a). A band inversion at $\G$ at fixed filling will now result in the band structure shown in Figure \ref{WeylAtPnt}(b), which exhibits a node at the high symmetry point $\G$.

The dispersion near the node can again be found by considering the tensor product representation of the two intersecting bands. In this case, we find that the dispersion cannot be linear, because there is no vector representation in the direct sum decomposition of the tensor product. Heuristically stated, the inversion at $\G$ imposes too many constraints on a general Hamiltonian near $\G$ for it to contain any linear terms. 

The semimetallic phase can again be characterised using the topological indices $N_{\pm}$. For the example in Figure \ref{WeylAtPnt} we find 
\bea
N_{-}^{l_1,l_3} &=& 0\notag \\
 N_{+}^{l_1} = -N_{+}^{l_3} &=& 1.
\eea
In contrast to \eqref{Weylcharac}, the characterisation this time involves two high symmetry lines. This can be understood by noting that the inverted band at $\G$ connects to the even representations in the set of valence bands at both $X$ and $M$. For crystals with space group $p4mm$, generic inversions at either $M$ or $\G$ can in fact always be characterised by a relation of the type
\be
N_{r}^{l_i} = - N_{r'}^{l_j}
\ee
Here $l_i \neq l_j$ are the two high symmetry lines adjacent to either $M$ or $\G$, and the indices $r$ and $r'$ are not necessarily the same. That is, for $p4mm$ there are eight possible sets of integers characterizing eight distinct semimetallic phases with a node at a high symmetry point.

\subsection{Two doubly degenerate bands}
\begin{figure}
\centering
\def\svgwidth{6cm}
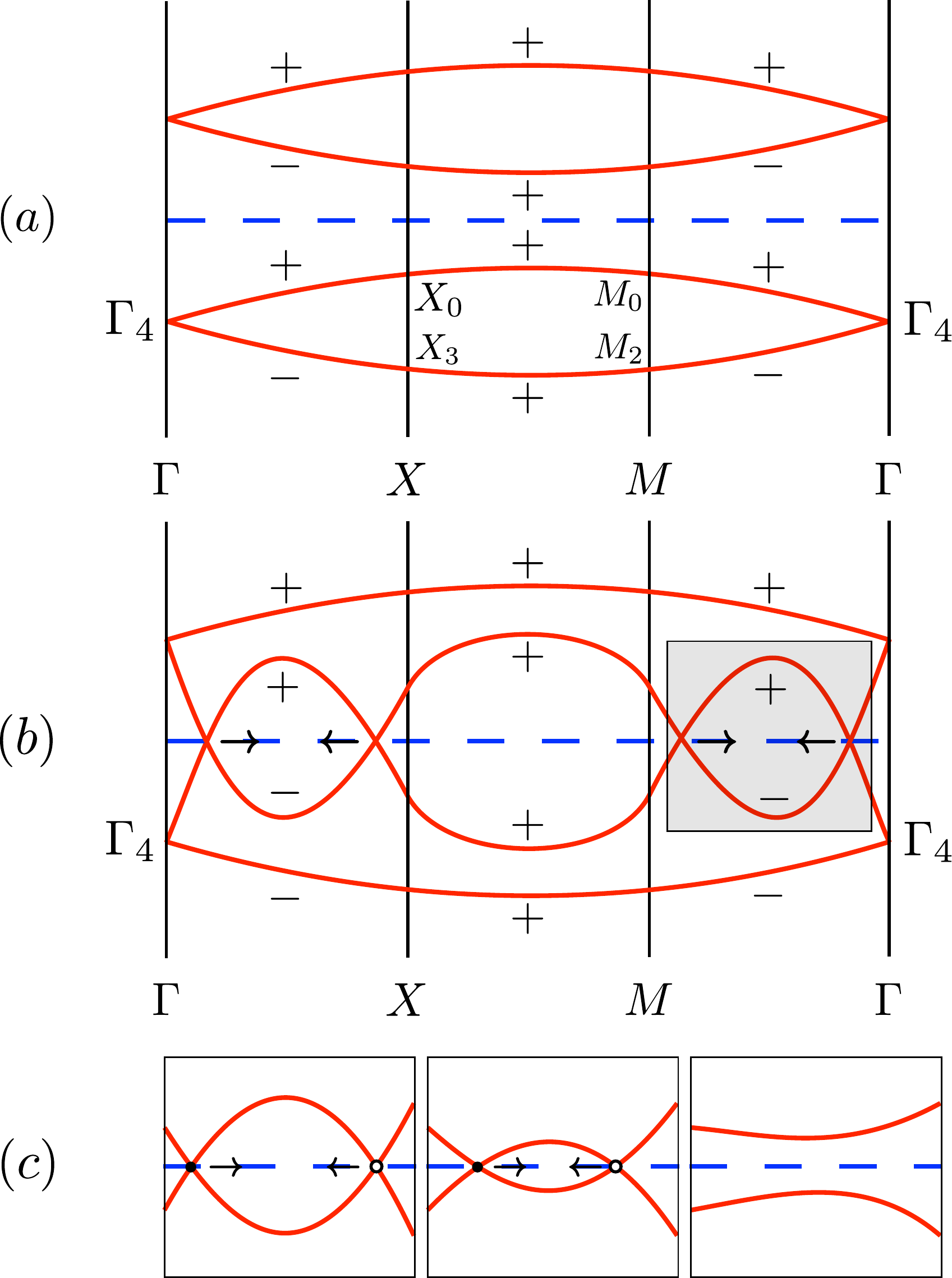
\caption{\label{doublebands}Inverting two doubly degenerate bands. (a) A topologically insulating state with two valance bands and two conduction bands, which are both degenerate at $\G$. (b) After a band inversion at $\G$, there will be four band crossings on each high symmetry line. Two of them will be avoided and not drawn, whereas the other two constitute Weyl nodes. (c) Zooming in on the boxed region in (b), the two opposite Weyl nodes along a single high symmetry line can be moved towards each other and annihilated. A direct transition between the original and final topologically insulating states is therefore possible. Notice that the final state is the same as the original one. This agrees with the fact that the integers $N_{+}$ and $N_{-}$ are all zero in both the initial and final configurations.}
\end{figure}

The final possibility within the $p4mm$ setting, is a scenario with two conduction and two valance bands, which are both doubly degenerate at a high symmetry point. Figure \ref{doublebands} shows this situation with the degeneracies occurring at $\G$. As we will see, this case contains an example of a direct transition between distinct topologically insulating phases, which circumvents any intermediate semimetallic phase. Suppose both the conduction and valance bands are in the irreducible representation $\G_4$ at $\G$. Performing an inversion at $\G$ will not change the topological phase according to the classification in terms of the indices $n_i^{\mathbf{k}}$, because the inversion does not change $n_4^{\G}$. The same conclusion can be reached by considering the band structure directly, starting from the situation in Figure \ref{doublebands}(a). The inversion will result in four band crossings each along both $l_1$ and $l_2$. Two of these crossings on each line are avoided, and two constitute actual nodes. The resulting band structure is thus as shown in Figure \ref{doublebands}(b). The two nodes along a given high symmetry line however, can be mutually annihilated. During such a process, a gap is opened up again, as shown in Figure \ref{doublebands}(c). This means that only a topologically insulating phase is stable after the inversion at the high symmetry point, and a transition between the two insulating states \textit{can} therefore happen directly, in contrast to the cases discussed before. Notice that in this case the initial and final topological insulators have the same integers $n_{i}^{\mathbf{k}}$ and are thus the same phase. A genuine direct transition between different phases can be obtained when the two double degenerate bands are different.

\begin{table*}
\begin{tabular}{c|ccccccccc}
WpG & $p1$ & $p2$ & $pm$ & $pg$ & $cm$ & $p2mm$ & $p2mg$ & $p2gg$ & $c2mm$  \\[0.1cm]\hline
Direct transition & $\G$ & $\G,X,M,Y$ & $\times$ & $\times$ & $\times$ & $\times$ & $\times$ & $\times$ & $\times$ \\
Node on line & $\times$ & $\times$ & $\times$ & $\times$ & $\times$ & $\G, X, M, Y$ & $\G, Y$ & $\G, M$ & $\G, Y'$\\
Node on point & $\times$ & $\times$ & $\times$ & $\times$ & $\times$ & $\times$ & $\times$ & $\times$ & $\times$  \\[0.3cm]
WpG & $p4$ & $p4mm$ & $p4gm$ & $p3$ & $p3m1$ & $p31m$ & $p6$ & $p6mm$ \\[0.1cm]\hline
Direct transition &  $\G, X, M$ & $\times$ & $\times$ & $\G, M, M'$ & $\times$ & $K$ & $\G, K, K'$ & $\G$ \\
Node on line & $\times$ & $\G, X, M$ & $\G, M$ & $\times$ & $\G, K, K'$ & $\G, X, Y$ & $\times$ & $\G, K, X$\\
Node on point & $\times$ & $\G, M$ & $\G, M$ & $\times$ & $\G\star, K\star, K'\star$ & $\G \star $ & $\times$ & $\G, K \star$ 
\end{tabular}
\caption{\label{transitions} Summary of all possible phases resulting from band inversions within two-dimensional topological insulators in class $A$. Direct transitions between two distinct topologically insulating phases may result from band inversions at the listed high symmetry points in each wallpaper group (WpG). If a stable intermediate phase results from an inversion, it contains either single Weyl node(s) along high symmetry lines, or a Weyl node at a high symmetry point. The points indicated in the former case are the high symmetry points at which a band inversion results in a node along a connecting high symmetry line. The dispersions near these nodes are always linear. For the latter case, the listed points indicate both the place of the band inversions, and the locations of the resulting nodes. The dispersions near these nodes is linear only in the instances marked by $\star$. Standard notation has been used to indicate the locations of high symmetry points, and a cross indicates that no transition of the listed type is possible.
%The $(-)$ conveyes that after an inversion only the less stable intermediate phase with two nodes along a high symmetry line in the fundamental domain is formed. 
%All high symmetry points are indicated by their familiar letters, $\G = (0,0)$, $X = (\pi,0)$, $M = (\pi,\pi)$, $Y = (0,\pi)$, $M' = (2\pi/3,2\pi/3)$ and $K = (2\pi/3,-2\pi/3)$. 
%The bar refers to negative instance of the coordinate, i.e $X=-\bar{X}$.
}
\end{table*}

\subsection{General wallpaper group}
The analysis of intermediate phases in crystals with $p4mm$ symmetry can straightforwardly be generalised to other wallpaper groups. As before, we will only consider inversions at high symmetry points. 

We start with the simplest wallpaper groups, $p1$, $p2$, $p3$, $p4$ and $p6$, which have $\mathbf{Z}_n$ as their point groups. In contrast to $p4mm$, these wallpaper groups do not any contain reflection-symmetric (high symmetry) lines. Since an inversion in that case cannot result in bands with different reflection eigenvalues crossing each other, the inversion will not result in the formation of a node. That is, these systems always remain gapped after an inversion, and we find only direct transitions between topologically insulating phases, without any intermediate phases. The same conclusion can be reached by considering the relations between the $n_i^{\mathbf{k}}$, which in this case only constrain the total number of bands at each high symmetry point to be equal. In fact, whenever a general space group gives rise to isolated high symmetry points, which are not connected to any high symmetry lines, direct transitions are an inevitable consequence of inversions at that isolated point. For the wallpaper groups, this situation occurs only for $p31m$, which thus allows for direct transitions whenever a band inversion takes place at the $K$ point. 

Direct transitions may also occur when a band inversion interchanges two different doubly degenerate sets of bands, as shown in Figure \ref{doublebands}. The only wallpaper group that has two different two-dimensional irreducible representations, and thus allows for this type of direct transition, is $p6mm$.

For wallpaper groups with a reflection symmetry, the nodes are constrained to lie on the reflection-invariant line, as in the case of $p4mm$. Again, having two nodes along a single line can be considered to be an unstable situation, since the nodes may move towards each other and annihilate. More stable single nodes along a high symmetry line cannot occur for wallpaper groups $pm$, $pg$, and $cm$, which have $\mathbf{Z}_2$ as their point groups. These structures contain high symmetry lines but {\it no} high symmetry points. Consequently, an inversion only creates pairs of nodes along high symmetry lines. For wallpaper groups with point groups other than $\mathbf{Z}_2$, single nodes along high symmetry lines can occur.

Besides nodes along high symmetry lines, Weyl nodes can also come about at high symmetry points. For this to happen, the inversion has to interchange a non-degenerate band with a doubly-degenerate band at fixed filling, as shown in Figure \ref{WeylAtPnt}. In all cases where a node appears either on a high symmetry line or on a high symmetry point, it can be checked whether or not the resulting dispersion around the node is linear. As before, this is done by evaluating the direct sum decomposition of the tensor representation of the two intersecting bands. We find that the dispersions for the nodes along reflection lines are always linear, while for most nodes at high symmetry points they are not. The only exception to the latter rule, are points with a $D_3$ symmetry, such as the $K$ point in graphene \cite{Neto2009}. 

The intermediate phases in general wallpaper groups can again be characterized by considering whether the relations between topological indices $n_i^{\mathbf{k}}$ are satisfied or not. Mismatches in these relations signal the presence of intermediate phases with gapless excitations. To define such mismatches more precisely, we take differences of the relations between high symmetry points. For the even ($+$) or odd ($-$) bands along a high symmetry line $L$, we get a single relation, analogous to what was discussed in Section \ref{sec::genwpg}. The mismatch is then described by integers $N_{\pm}^L$, where a consistent orientation of all high symmetry lines is chosen in order to make $N_{\pm}^L$ well-defined. For example, for $p4mm$, this means that we define 
\be
N^{l_1}_+ = n_0^{\G} + n_2^{\G} + n_4^{\G} - n_0^X - n_2^X
\ee
whereas $N^{l_3}_+$ is defined as
\be
N^{l_3}_+ = n_0^{M} + n_3^{M} + n_4^{M} - n_0^{\G} - n_3^{\G} + n_4^{\G}.
\ee
A single node along a high symmetry line $L$, results in a mismatch given by
\be
N_{\pm}^L = - N_{\mp}^L, \quad |N_{\pm}^L| = 1.
\ee
In the case of a node on a high symmetry point between two high symmetry lines $L$ and $L'$, the characterisation is similar:. 
\bea
N^{L}_{+} = N^{L'}_{+} &=&0 \notag \\
 N^{L}_{-} = - N^{L'}_{-}, 
\eea
or similarly with $+$ and $-$ interchanged. In this expression, $|N_{\pm}^{L,L'}| = 1$ and $L \neq L'$. This concludes the analysis of all possible transitions upon a band inversion respecting a wallpaper group symmetry. The allowed intermediate phases and direct transitions for each wallpaper group and for all possible inversions at high symmetry points are listed in Table \ref{transitions}. The table also indicates whether or not the dispersion around a node can be linear.

\subsection{Nodes without fine-tuning}
The transitions between phases we discussed so far generically require fine-tuning of some of the parameters in the Hamiltonian, i.e. to achieve the band inversion we needed to tune a conduction band below the valance band. However, it is also possible to have nodes without fine-tuning that are stable under small deformations of the Hamiltonian which preserve the crystal symmetry.
%\textcolor{red}{In general, one may expect that creating a Weyl node in a given material requires fine-tuning of some of the parameters in the Hamiltonian. The nodes emerging in the transitions between topological phases discussed above, however, are stable under small deformations of the Hamiltonian preserving the crystal symmetry.}
The location of the node in reciprocal space may be changed by the deformation, but for a sufficiently small deformations it will not disappear. To understand why these nodes are stable, one can simply enumerate the different constraints allowed on the parameters in a local Hamiltonian around nodes in two and three dimensions. In doing so, we take into account the local band structure around the node only, and ignore possible global constraints on the representations.

As before, we consider high symmetry locations $M$, which can be points, lines, or planes in the Brillouin zone that are left invariant by a little co-group $G^M$. For simplicity, we restrict attention to symmorphic space group symmetries here. the number of bands in a given irreducible representation $M_j$ of the group $G^M$ is given by the integer $n_j^M$. In general, the energy eigenvalues for all bands are different, but there can be special points where eigenvalues coincide. This requires a fine-tuning of $N= \sum_j (n_j^M)^2 -1$ parameters in the Hamiltonian. Locally stable degenerate points on $M$ can exist if the dimension of $M$ is greater than or equal to $N$. A well-known example of this argument in three dimensions is to consider the entire Brillouin zone without imposing any symmetry constraints. The group $G^M$ is then the trivial group, containing only the identity. It only has a trivial representation $M_1$, so that with two bands in the Brillouin zone we must have $n_1^M=2$ and $N = (n_1^M)^2-1=3$. It is thus possible to have locally stable degeneracies in the bulk of a three-dimensional Brillouin zone without any symmetries. These are the well-known Weyl nodes of a Weyl semimetal. 

Applying the same arguments to materials in two dimensions, it is clear they cannot contain stable degenerate points at generic points in the Brillouin zone. Along high symmetry lines in two dimensions however, there always is a $\mathbf{Z}_2$ symmetry which gives rise to two representations, an even and an odd one. If we have one band in each representation this means $N=1$, only one parameter needs to be tuned in order to let the two bands become degenerate. This can therefore happen at isolated points on the high symmetry line. high symmetry points in two dimensions may be left invariant by all kinds of different symmetry operations, but since a high symmetry point is only a single location in reciprocal space, there is no room to fine-tune the location of any node, and representations generically do not become degenerate.

In three dimensions, besides Weyl nodes in the bulk of the Brillouin zone, we should also consider isolated high symmetry points, lines and planes. The case of a plane is similar to that of a line in two dimensions. The plane is always left invariant by a $\mathbf{Z}_2$ symmetry operation, and an even and odd band can become degenerate. We only need to tune one parameter to achieve this, and one can therefore have an entire curve of degenerate points in the high symmetry plane. Along a high symmetry line in three dimensions more interesting representations can appear, and two distinct irreducible representations can become degenerate at locally stable points. Finally, as in two dimensions, different representations are generically non-degenerate at high symmetry points.

The possibilities considered so far describe bands becoming degenerate on a high symmetry location $M$ left invariant by the group $G^M$. A separate possibility arises if we consider bands which become degenerate at special points $H$ within $M$ at which the group $G^M$ is enhanced to a larger group $G^H$, thus forcing several representations of $G^M$ to combine into a single irreducible representation of $G^H$. This is precisely what happens for example when two bands along a high symmetry line connect to a high symmetry endpoint of that line. These types of connections are the key ingredient in the combinatorial arguments presented in the previous sections. When looking for possible nodes, we should therefore also consider single higher-dimensional representations at high symmetry points or lines, and examine how they split when going away from that high symmetry location.

To see whether there is a node when bands become degenerate, we need to consider the behavior of the Hamiltonian in a small neighborhood of the degeneracy. If $G^H$ is the little co-group at $H$, the degenerate bands from some degenerate representation $H_j$ of $G^H$. The momenta in the directions perpendicular to $H$ will also transform in a particular way under the symmetry operations of $G^H$, and will form a corresponding representation $V_i^H$. Terms in
the Hamiltonian which are linear in momentum, are only allowed if a vector representation $V_i^H$ appears in the direct sum decomposition of the tensor product $H_j \otimes H_j^{\ast}$, because the Hamiltonian itself must be invariant under $G^H$. One can easily check in explicit examples that whenever this happens, there will be a node: the energy eigenvalues depend linearly on momentum, and are non-analytic at the degenerate locus. It is beyond the scope of this paper to attempt to prove this in full generality. 

The list of all possible locally stable nodes in two and three dimensions can be summarised as follows. For generic points in two dimensions, there are no locally stable nodes. For high symmetry lines in two dimensions, isolated nodes along the line are allowed at places where an even and an odd band meet. High symmetry points in two dimensions will feature nodes if they have $D_3$ symmetry, and there is a band transforming in the two-dimensional vector representation of $D_3$. In three dimensions, Weyl nodes are possible at generic points, regardless of the crystal symmetry. On high symmetry planes in three dimensions, a curve of nodes may appear as an even and an odd band become degenerate. Similarly, a line of nodes may appear along a high symmetry line in three dimensions, if there is a band in an irreducible representation $M_j$ such that $M_j \otimes M_j^{\ast}$ contains a vector representation $V^M$. An isolated node may appear along a high symmetry line in three dimensions, if there are bands in two irreducible representations $M_1$ and $M_2$ such that $M_1\otimes M_2^{\ast}$ contains a vector representation $V^M$. Finally, a high symmetry point in three dimensions may host a node if there is a band in an irreducible representation $M_j$ such that $M_j\otimes M_j^{\ast}$ contains a vector representation $V^M$. A detailed analysis of the degeneracies and their dispersion in three dimensions with time-reversal symmetry is in fact discussed in \cite{Bradlynaaf5037}.

\section{Generalizations \label{sec::gen}}
\begin{figure}
\centering
\def\svgwidth{5cm}
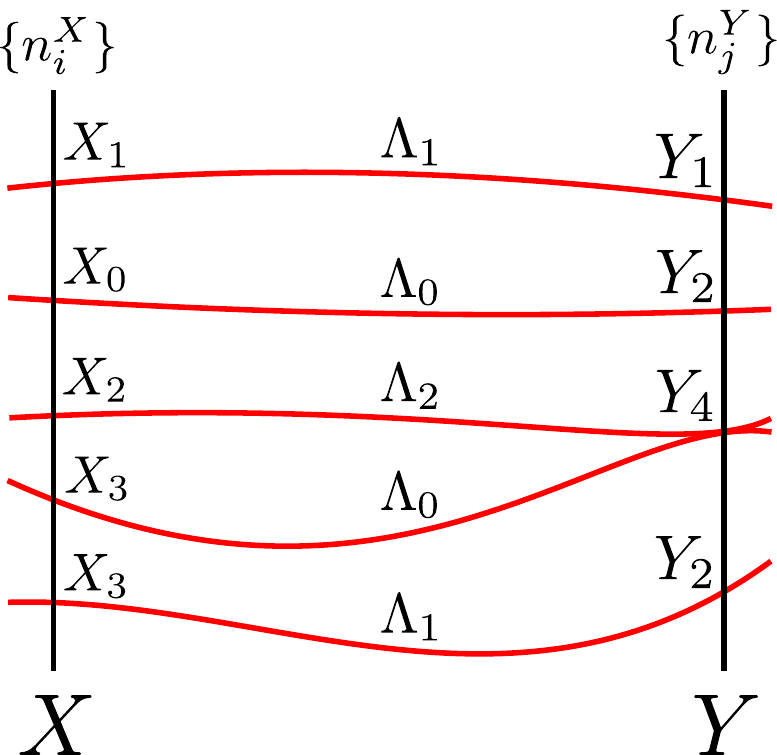
\caption{\label{argumentinpic} Sketch of the relations between high symmetry points imposed by high symmetry lines. A set of valance bands is shown between two generic high symmetry points $X$ and $Y$, which are left invariant under the symmetry transformations contained in the little co-groups $G^X$ and $G^Y$. The irreducible representations of $G^X$ and $G^Y$ may be labelled as $X_i$ and $X_j$. Each of the bands needs to fall within one of these representations at the corresponding high symmetry points. The representations on $X$ and $Y$, however, need to be compatible with the representations $\Lambda_j$ of the high symmetry line connecting $X$ and $Y$. The remaining set of independent integers indicating how many bands are in which representations at the high symmetry points, finally determines the number of possible topologically distinct configurations of the set valence bands, matching the abstract K-theory classification.}
\end{figure}

The combinatorial arguments leading to a classification of topological phases in class $A$, and a characterisation of phases resulting from band inversions within such two-dimensional topological insulators, can be readily extended to three-dimensional crystal structures. For a general space group $G$ and its associated first Brillouin zone with high symmetry points $\mathcal{M}^i$, the first step is to determine the representations of the little co-groups $G^{\mathcal{M}^i}$ for all $\mathcal{M}^i$. A set of integers $n_j^{\mathcal{M}^i}$ can then be introduced to indicate the number of valence bands in representation $j$ at high symmetry point $\mathcal{M}^i$. These integers are not independent, because they are constrained by the compatibility relations imposed by high symmetry lines connecting various $\mathcal{M}_i$, as shown pictorially in Figure \ref{argumentinpic}. Giving a list of values for a complete set of independent integers $n_j^{\mathcal{M}^i}$ amounts to a characterisation of the set of valence bands, so that finding the number of independent integers in a given space group is equivalent to classifying its possible topological phases. In three dimensions, it is possible to have high symmetry planes, but these do not add any compatibility relations beyond those already imposed by the high symmetry lines.

As in the two-dimensional case, the combinatorial argument does not indicate the possible values of Chern-numbers, which need to be included in a full classification of topologically distinct phases. Fortunately, they can be obtained in a straightforward manner. Chern numbers are always given by two-dimensional integrals. For example, the TKNN invariant is an integral over the full Brillouin zone in two dimensions. For crystals in three dimensions, the Chern number will likewise be given by an integral over a two-dimensional plane within the first Brillouin zone. In the absence of band crossings, the Chern number can be evaluated as a sum of contributions from the integration of individual valance bands. The Chern numbers evaluated on two parallel two-dimensional planes must be equal by continuity. 

If a three-dimensional crystal contains high symmetry planes, these may be used as convenient locations for defining a set of Chern numbers. Firstly, such a plane can host a nonzero Chern number if and only if there is no reflection symmetry within the plane, mimicking two dimensional case. Moreover, a high symmetry plane in three dimensions is always left invariant by a reflection acting perpendicular to the plane, under which the bands can be even or odd. Separate Chern numbers $c_{\pm}$ can then be assigned to the full set of even or odd bands, and are obtained by summing the contributions from individual even or odd bands. For a full characterisation of the topological phase, both numbers $c_{+}$ and $c_{-}$ need thus to be specified for all high symmetry planes in the first Brillouin zone. However, these Chern numbers are not independent in the same sprit as the constraints found above. Namely, a general plane in the Brillouin zone, on a small distance away from the high symmetry plane, may have its (single) Chern number $c$ equal to zero. In that case, $c_+$ must equal $-c_-$ on the high symmetry plane, to ensure continuity. Therefore only a nontrivial {\it mirror} Chern number $c_m = (c_+ - c_-)/2$ can be defined in this scenario \cite{Teo2008}. 

The other possibility is that the general plane has a non-zero Chern number $c$, so that $c_+ + c_-$ must equal $c$ by continuity. This situation results in relations between the Chern numbers on distinct but parallel high symmetry planes. For example, suppose there are mirror planes at $k_z = 0$ and $k_z = \pi$ that have Chern numbers $c_{\pm}^0$ and $c_{\pm}^{\pi}$, respectively. If a general plane between $k_z = 0$ and $k_z=\pi$ has its total Chern number equal to $c$, then this implies
\bea
c &=& c_+^0 + c_-^0 \notag \\
c &=& c_+^{\pi} + c_-^{\pi}.
\eea
Combining these two equations then yields the relation
\be
c_+^0 + c_-^0 = c_+^{\pi} + c_-^{\pi}.
\ee
Out of the four Chern numbers characterising the two high symmetry planes, only three are independent. Notice that these are exactly the same type of relations as the constraints between high symmetry points that we introduced in Section \ref{sec::genwpg}. The combination of all independent Chern numbers and the set of independent integers obtained from the representations of the valance bands, completely specify the topological phases in class $A$. In Appendix \ref{sec::Mat} we further argue that other topological invariants are not present for class $A$ and our arguments indeed provide a complete classification.

\begin{figure}
\centering
\def\svgwidth{4cm}
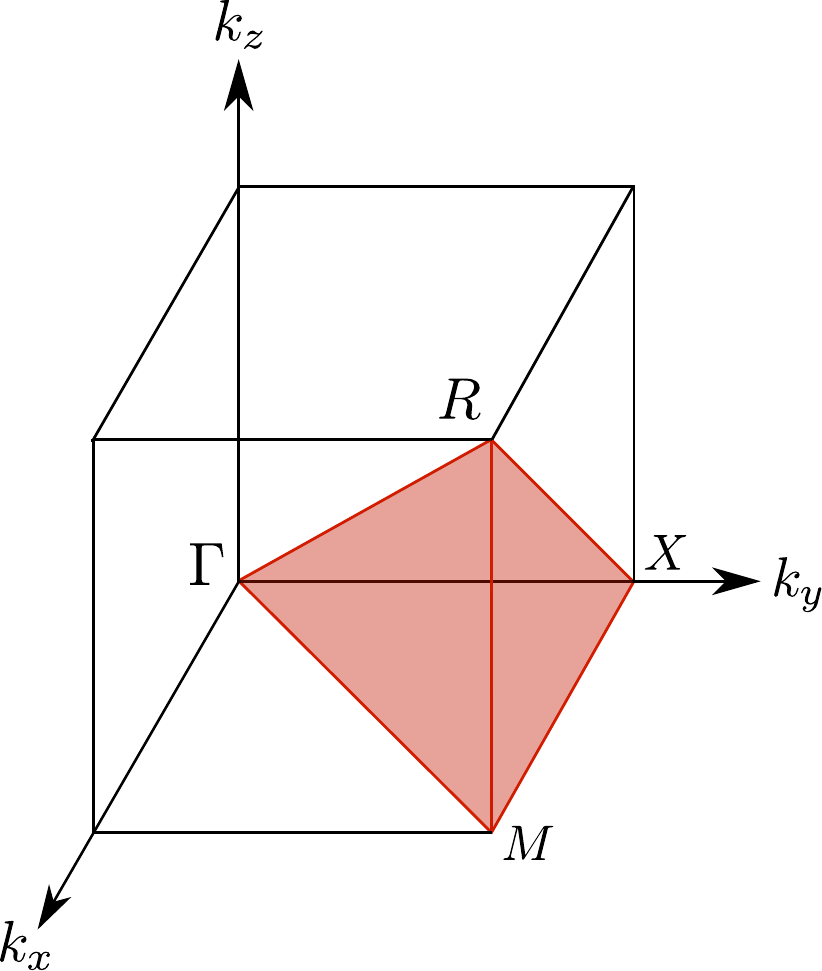
\caption{\label{FundOh} Fundamental domain of the octahedral group $O_h$ (shaded red). The red high symmetry lines connect the high symmetry points $\G$, $M$, $X$ and $R$.}
\end{figure}

As a concrete working example of the classification scheme in three dimensions, consider the symmorphic space group $Pm\bar{3}m$ (no. 221). It has an octahedral point group, $O_h$, which contains the symmetries of a cube and has $48$ elements. $O_h$ is generated by the following elements
\bea
r\cdot k &=& (-k_z,k_y,k_x), \notag \\
t\cdot k &=& (k_y, k_x, -k_z), \notag \\
I\cdot k &=& -k. 
\eea
The fundamental domain is shown in Figure \ref{FundOh}. It contains six high symmetry lines coming from the two, three and fourfold rotation axes. The endpoints of these lines are the high symmetry points $\G$, $R$, $X$ and $M$, which have little co-groups $O_h$, $O_h$, $D_4 \times \mathbf{Z}_2$ and $D_4 \times \mathbf{Z}_2$, respectively. The high symmetry points all have $10$ different irreducible representations, so that $40$ integers can be introduced to specify the set of valence bands. The high symmetry lines yield $25$ relations between these integers, which in this particular case were in fact already found by Wigner \cite{Wigner1936}. Taking into account the dependencies between the relations, $22$ integers remain to be specified in order to fully characterise the set of valence bands. The relations coming from mirror planes do not add any additional constraints on these $22$ integers, since they are already implicitly satisfied by taking into account the constraints coming from the high symmetry lines. 

The octahedral point group has no Chern numbers, because within each high symmetry plane there is a perpendicular reflection plane. Hence, we find that a topological phase in class $A$ protected by $O_h$ symmetry is classified by $\mathbf{Z}^{22}$, i.e  by $22$ integers. Once again this result can be corroborated by calculations from a more formal, mathematical perspective. As shown in Appendix \ref{sec::Mat}, the result from such a $K$-theory calculation is exactly the same as that obtained in our combinatorial approach. 

Having found that the combinatorial arguments can be applied to three dimensional space groups in essentially the same way as in two dimensions, the question arises if it can also be extended to systems beyond class $A$, which include time reversal, particle-hole, or chiral symmetries. In fact, the two-dimensional symmorphic space groups in class $AIII$ can directly be addressed, using the complete classification of topological matter in class $A$. This can be done because there exists an isomorphism relating the $K$-theory computation for class $A$ in three dimensions to two pieces in two dimensions, one of which corresponds to class $A$ while the other corresponds to class $AIII$. We are not aware, however, of a similar mapping for nonsymmorphic space groups, and so refrain from a detailed discussion in the present work. 

In the case of class $AII$, of time-reversal symmetric topological insulators, the time-reversal symmetry and crystal symmetry intricately intertwine, which is anticipated to give a richer structure than the one presented here. Nevertheless, we anticipate that the key ideas identified in the combinatorial approach hold also in that class, and may provide new insights. Indeed, relations between high symmetry points in the Brillouin zone for class $AII$ materials underly to the classification procedure of Ref. \cite{Slager2013}. These notions then reduce to the weak invariants if the space group symmetries are subsequently neglected, as time reversal symmetry acts invariantly on high symmetry planes. Applying the types of arguments presented in the present paper to representations of magnetic point groups may be a way to further characterise these types of topological materials.

\section{Conclusions}
In this paper, we presented a straightforward combinatorial procedure that can be used to classify all distinct topological phases within class $A$, by taking into account the space group symmetries of a material. Although the arguments presented involve only basic representation theory applied to the electronic band structure, we can be confident that the resulting classification is complete, since it agrees with the formal, but mathematically involved, $K$-theory computation \cite{Freed2013, RoyerFreed}. Indeed, besides providing an efficient and easily applicable algorithm for determining the number of possible topological phases protected by a given space group symmetry, the combinatorial arguments also provide physical insight in the distinctions between topological phases and the according mathematical framework. This may in itself be a useful starting point for future research, for example by providing a possible route towards new $K$-theoretical computations in systems beyond the scope of the current work. 
 
Within the context of materials in class $A$, the formalism introduced here provides a way of identifying possible transitions between topological phases as well as the phases themselves. In particular, it allows us to list all possible gapless phases at the boundaries between distinct topological phases for a given space group. We use this to explicitly map all possible Weyl phases in two-dimensional materials in class $A$, and provide general criteria with respect to their stability. 

Interestingly, our results can also be applied in the context of high energy physics, in particular in string theory. That is, D-branes in string theory carry charges which are classified using $K$-theory \cite{Witten:1998cd}. For Type IIB superstrings on toroidal orbifolds, the setup is then exactly the same as  we have been discussing here. Our methods therefore also provide a simple way of counting D-brane charges in these types of superstring theories. 

Besides extending the current work beyond class $A$ and to other fields of physics, an interesting direction for future research is the application of the combinatorial arguments presented here to evaluate the symmetries that are left invariant at the physical boundaries of a piece of material. These types of arguments may then be used to identify and classify possible surface states, which are the hallmark of topological bulk order, and which would naturally complement the complete classification of class $A$ materials with a given space group, presented here.

~\\
{\it Acknowledgements} --- We would like to thank Wolfgang L\"{u}ck, Daniel Freed and Aaron Royer for illuminating discussions. CLK and RJS would like to thank the Instituut Lorentz at the University of Leiden where this work was initiated. CLK is supported by a Simons Investigator grant from the Simons Foundation. JvW acknowledges support from a VIDI grant financed by the Netherlands Organization for Scientific Research (NWO). JK is supported by the Delta ITP consortium, a program of the Netherlands Organisation for Scientific Research (NWO) that is funded by the Dutch Ministry of Education, Culture and Science (OCW).

\appendix

\section{Nonsymmorphic space groups \label{sec::NSyms}}
The combinatorial arguments presented in the main text can also be applied to nonsymmorphic symmetry groups. In contrast to the symmorphic case, however, we then need to explicitly take into account the translations as well as the point group contributions to the full space group. The translational part of nonsymmorphic groups gives rise to additional phase factors in the character tables which characterise the possible representations at high symmetry points. Besides having to include these phase factors, the method is the same as presented for symmorphic groups.

We will closely follow \cite{bradleycracknell} in our analysis of how having a nonsymmorphic space group symmetry affects the representations at fixed points in $\W$. In Seitz notation, nonsymmorphic elements are represented by $S_i = \{R_i|\mathbf{w}_i\}$, where $R_i$ is an element of the point group and $\mathbf{w}_i$ a fractional lattice translation. As an example of such a nonsymmorphic group element, consider a glide reflection $t_g$. If $(x,y)$ is a lattice point, then the glide reflection acts as
\be
t_g\cdot (x,y) = (x+1/2,-y).
\ee
This transformation has the property that $t_g^2 = (1,0)$, i.e. a pure lattice translation. In general, nonsymmorphic elements make it impossible to separate reflections and rotations from lattice translations. However, we can still use the action of just the point group contributions, i.e. the elements $R_i$, to determine the high symmetry locations in the first Brillouin zone. 

Suppose $\mathcal{C}$ is a set of wave vectors $\mathbf{k}$ held fixed by a little co-group $G^{\mathcal{C}}$. If this little co-group stems from a nonsymmorphic space group, some of its elements $S_i$ may themselves contain translations, making the representation theory of $G^{\mathcal{C}}$ much richer than what we saw so far. The translations in $S_i$ are fractional and constrain the representation $\rho^{\mathbf{k}}$ of $G^{\mathcal{C}}$ to satisfy
\be\label{projectrep}
\rho^{\mathbf{k}}(R_i)\rho^{\mathbf{k}}(R_j) = \exp\left(-i\mathbf{g}_i\cdot \mathbf{w}_j\right)\rho^{\mathbf{k}}(R_iR_j),
\ee 
Note that for symmorphic crystals, $\mathbf{w}_j = 0$. The reciprocal lattice vector $\mathbf{g}_i$ is defined by semimetal
\be
R_i^{-1}\mathbf{k} = \mathbf{k} + \mathbf{g}_i
\ee
Representations that satisfy \eqref{projectrep} are called projective representations. The bands on the high symmetry location $\mathcal{C}$  thus transform under projective irreducible representations rather than ordinary representations. 

In direct analogy to the procedure for symmorphic space groups, we need to determine the projective irreducible representations for the little co-groups in the nonsymmorphic case, count them, and finally find the relations between them that are imposed by the representations along high symmetry lines. The only difference between the treatments of symmorphic and nonsymmorphic symmetries in class $A$, is thus the type of representations used. 

As a concrete example, consider the wallpaper group $p2gg$. The associated point group is generated by two elements; one reflection $t_x$ in the $k_x$-direction and an inversion $\s$. Modulo lattice translations, this wallpaper group has the following elements,
\be
G/\mathbf{Z}^2 \simeq G_0 = \left\{\{e|00\},\{\s|00\},\{t_x|\tfrac{1}{2}\tfrac{1}{2}\},\{t_y|\tfrac{1}{2}\tfrac{1}{2}\}\right\},
\ee
where we denoted the identity element by $e$. Using the elements $\s$, $t_x$, and $t_y$, we find that there are four fixed points, $\G$, $X$, $M$, and $Y$, which all have little co-group $G_0$. There are also four lines that are held fixed by reflections in the two axes. These lines connect the four high symmetry points and form the boundary of the fundamental domain $\W$. These results are summarised in table \ref{fixedp2mm}.
\begin{table}
\centering
\begin{tabular}{c|c}
$\mathcal{C}^n_i$ & \rm{Stabilizer group}\\\hline
$\G = (0,0)$ & $\mathbf{Z}_2\times\mathbf{Z}_2 = \{1,\s , t_x , t_y\}$\\
$X = (\pi,0)$ & $\mathbf{Z}_2\times\mathbf{Z}_2 = \{1,\s , t_x , t_y\}$\\
$M = (\pi,\pi)$ & $\mathbf{Z}_2\times\mathbf{Z}_2 = \{1,\s , t_x , t_y\}$\\
$Y = (0,\pi)$ & $\mathbf{Z}_2\times\mathbf{Z}_2 = \{1,\s , t_x , t_y\}$\\
$l_1 = (\a,0)$ & $\mathbf{Z}_2 = \{1,t_y\}$\\
$l_2 = (\a,\pi)$ & $\mathbf{Z}_2 = \{1,t_y\}$\\
$l_3 = (0,\a)$ & $\mathbf{Z}_2 = \{1,t_x\}$\\
$l_4 = (\pi,\a)$ & $\mathbf{Z}_2 = \{1,t_x\}$
\end{tabular}
\caption{\label{fixedp2mm} The high symmetry locations for $p2gg$. The same structure is also valid $p2mm$ and $p2gm$. The corresponding fundamental domain $\W$ is given by the first quadrant of the first Brillouin zone, i.e. $0\leq k_{x,y} \leq \pi$.}
\end{table}

The nonsymmorphic elements do not enhance the representations at $\G$, because $\mathbf{g}_i = 0$ for all elements $R_i$. There could however, be projective representations at $X$, $Y$, and $M$, as used for example in the analysis of Ref. \cite{Young2015}. First, consider the high symmetry point $X$, and define the object $\n(R_i,R_j)$ as:
\be
\n(R_i,R_j) = \exp\left(-i\mathbf{g}_i\cdot \mathbf{w}_j\right),
\ee
In this case, all $\n$'s are unity except for $\n(t_x,t_x) = \n(t_x,t_y) = \n(\s,t_x) = \n(\s,t_y) = -1$. This can easily be seen by noting that $g_e = 0$, $g_{\s} = -g_1$, $g_{t_x} = -g_1$, and $g_{t_y}=0$, with $g_1 = (2\pi,0)$. This information is enough to determine the representations at $X$, using the standard theory of projective representations. We find that these values of $\n$ allow only for the two dimensional representation of $D_4$ to appear. Hence, at $X$ there is a single two dimensional representation, and all bands must be pairwise degenerate. 

Repeating the analysis for $Y$ and $M$, one easily verifies that the two dimensional representation of $D_4$ is found also at $Y$, whereas at $M$ there are four possible irreducible representations of the group $\mathbf{Z}_2\times\mathbf{Z}_4$. From here, we can again set up the combinatorial arguments in terms of the number of bands in each allowed representation at high symmetry points and the constraining relations between them. To define the constraints, we need to determine the representations along high symmetry lines. We do this by including the phase factors arising once the representations $\rho^{\mathbf{k}}$ of the little co-group are promoted to representations of the space group.  More precisely, the representations $\rho^{\mathbf{k}}$ are related to space group representations $D^{\mathbf{k}}$ by
\be
D^{\mathbf{k}}(\{R_i|\mathbf{w}_i\}) = \exp(-i\mathbf{k}\cdot\mathbf{w}_j)\rho^{\mathbf{k}}(R_i)
\ee
The phase factor $\xi = \exp(-i\mathbf{k}\cdot\mathbf{w}_j)$ appear in the character tables and are only present for nonsymmorphic elements. In the space group $p2gg$, we have $\mathbf{w}_i = (1/2,1/2)$. For $X$ and $Y$ this implies $\xi = -i$, while for $M$ it yields $\xi = -1$. Moreover, we have $\xi = e^{-i\a/2}$ for $l_1$ and $l_3$, and $\xi = -ie^{-i\a/2}$ for $l_2$ and $l_4$. 

From here, we can assign an integer to each representation of the little co-group at high symmetry points. This gives $4$ integers at $\G$ and $M$ and one integer for both $X$ and $Y$, hence $10$ integers in total. There are also $7$ independent relations between these integers, giving rise to only $3$ independent integers. The representation theory of the set of valance bands for a crystal with $p2gg$ symmetry is thus specified by $3$ integers, in agreement with the $K$-theory computation. For the other two-dimensional nonsymmorphic space groups, $p2gm$, $p4gm$, and $pg$, the analysis is similar. The results can be found in table \ref{table2d}. 

The lowest symmetry group in the table, $pg$, is interesting because the nonsymmorphic elements in this case only affect high symmetry lines, on which new constraints emerge. Young and Kane \cite{Young2015} already noted these specific constraints rooted in the nonsymmorphic nature of the space group. The little co-group keeping the high symmetry lines in $pg$ fixed, is the factor group $G/\mathbf{Z}^2$, which consists of two elements:
\be
G/\mathbf{Z}^2 \simeq \{ \{e|00\}, \{t_y |\tfrac{1}{2}0\} \}.
\ee
Here $t_y$ is a reflection in the $k_x$ axis. For this factor group we obtain the character table shown in table \ref{pg}, where we defined $\xi(\a) = e^{-i\a/2}$. Because the fixed line $l = (\a,0)$ goes around the full Brillouin zone, the representation of the valance band must be periodic. This is only possible when the number of bands in $\rho_0$ equals that in $\rho_1$. Imposing a similar constraint along the other fixed line at $k_y = \pi$, and imposing that the total number of bands on each high symmetry line is equal, we find that there is a single integer specifying the topological phase of crystals in class $A$ with $pg$ symmetry. This again agrees with the $K$-theory computation as dicussed in appendix \ref{sec::Mat}.

\begin{table}[t]
\centering\vspace{0.5cm}
\begin{tabular}{c|ll}
& $\{e|00\}$ & $\{t_y|\tfrac{1}{2}0\}$ \\\hline
$\rho_0$ & $1$ & $\xi(\a)$\\
$\rho_1$ & $1$ & $-\xi(\a)$
\end{tabular}
\caption{\label{pg} Character table of $pg$ along $l_1 = (\a,0)$ and $l_2 = (\a,\pi)$ with $\xi(\a) = e^{-ia/2}$.}
\end{table}

\section{Mathematical details \label{sec::Mat}}
n this Appendix we will motivate some mathematical details that were omitted in the main text. Although we will not be completely rigorous, we give arguments from both a mathematical and physical point of view to substantiate our claims. We note that the main text is self-consistent and may be read independently from the following discussion. We will focus on the symmorphic symmetries for simplicity and refer to a detailed discussion on nonsymmorphic symmetries and their incorporation in K-theory to \cite{Freed2013, RoyerFreed}.

The mathematical classifications of gapped free fermion theories protected by symmetry groups, all stem from work by Horava in 2005 \cite{Horava2005}, who noticed a connection between Fermi surfaces and $K$-theory, which was further elaborated on by Kitaev in 2009 \cite{Kitaev2009}. In particular, Kitaev discussed gapped free fermions in various Altland-Zirnbauer (AZ) classes with discrete translational symmetry. These systems form the starting point for topological band theory which was recently shown to be relevant also to experimental setting. Nonetheless, this study failed to rigorously include the full crystal symmetry in its analysis. From the $K$-theory side, Freed and Moore attempted to fill this hiatus in 2013 \cite{Freed2013}. They pointed out what type of mathematical objects could classify topological phases in any AZ class, in the presence of arbitrary space group symmetry. 

The rest of the Appendix is organized as follows. In the section \ref{Appsetup} we discuss the general set-up and an the role of the translational symmetry in this regard. In particular we observe the emergence of  a vector bundle structure. This provides the basis for the next section in which we discuss the classification of these bundles using $K$-theory. Inclusion of the full space group symmetry is then discussed in section \ref{Appequivariant}. There, we will argue for a simple combinatorial way of computing the corresponding equivariant $K$-theory. At the end of this appendix, we will consider the space group $Pm\bar{3}m$ as an explicit example. Throughout this appendix, we do not use crystallographic terminology such as little co-groups or high symmetry locations, but rather the standard mathematical terminology such as stabilizer group and fixed point sets.   

\subsection{Set-up}
\label{Appsetup}
We are interested in the topological properties of class $A$ massive fermions on a $d$-dimensional lattice. These systems have a particular space group symmetry $\widehat{G}$. 

The dynamics of massive free fermions or insulator are governed by a gapped Hamiltonian $H$. Let $E$ be the eigenvalues of $H$ and $\ket{\psi}$ its eigenstates. We say that a Hamiltonian is gapped if there exists a range $|E| < \D$ for some $\D > 0$ such that $H$ does not have an eigenstate $\chi$ with eigenvalue $\a$ within this range in the infinite volume limit. As these gapped free fermions live on a lattice in $d$ dimensions, $H$ respects the lattice symmetry, i.e. $\rho(g)H = H\rho(g)$ with $\rho(g)$ a representation of the space group. Let us first consider the discrete translations in $\widehat{G}$. A lattice $\L$ is a subset of Euclidean space. It is isomorphic to $\mathbf{Z}^d$ and is spanned by orthogonal basis vectors $\mathbf{t}_i$, $i=1,\dots,d$. The discrete translation symmetry
\be
T_{\mathbf{n}}:\mathbf{v} \mapsto \mathbf{v} + n_i\mathbf{t}_i
\ee
with $\mathbf{n} = (n_1, \dots, n_d)$ in $\mathbf{Z}^d$ and $\mathbf{v}$ a lattice vector, constrain the fermions to form a representation of this symmetry. The representations are simple phases labelled by a $d$-dimensional momentum vector $\mathbf{k}$. More precisely, the representations are defined by
\be\label{reptrans}
\rho_{\mathbf{k}}(\{e|\mathbf{v}\}) = \exp(-i \mathbf{v}\cdot \mathbf{k}),
\ee
where we used Seitz notation to represent the discrete translation. This is basically a discrete Fourier transformation. The nature of these representations allows for a simple description of fermions in momentum space, because $k_i \sim k_i + g_i$ with $\{g_i\}$ a basis of momentum space such that $g_i\cdot g_j = 2\pi \delta_{ij}$. In momentum space, the fermions thus live on a $d$-dimensional torus; the Brillouin zone $\mathcal{M}$. The Brillouin zone is in fact the space of characters of the form given in \eqref{reptrans} and we will use $k$ as a parameterization. For each $\mathbf{k}$ vector we have a Hamiltonian $H(\mathbf{k})$ and Hilbert space $\mathcal{H}(\mathbf{k})$. The Hamiltonian $H(\mathbf{k})$ is related to the second quantized gapped Hamiltonian $H$ as
\be\label{hamsec}
H = \sum_{\mathbf{k}} H(\mathbf{k})c_{\mathbf{k}}^{\dagger}c_{\mathbf{k}}
\ee
where $c^{\dagger}_{\mathbf{k}}$ and $c_{\mathbf{k}}$ are creation and annihilation operators  such that $\{c_{\mathbf{k}}, c^{\dagger}_{\mathbf{k}'} \} = \delta(\mathbf{k}-\mathbf{k}')$. Besides discrete translational symmetry, lattices may also have reflection and rotation symmetries. For example, a square lattice in two dimensions has an extra $D_4$ symmetry coming from the symmetries of the square. These symmetries naturally act in position space, but also act in the Brillouin zone $\mathcal{M}$ in the transpose representation, as can be seen from equation \eqref{reptrans}. In momentum space, the states in $\mathcal{H}(k)$ will form a representation of the symmetry group that acts on $\mathcal{M}$. This symmetry group is the full space group $\widehat{G}$ modded out by discrete translations, and is denoted by $G$. More precisely, $\widehat{G}$ sits in the group extension 
\be\label{groupextension}
1 \to \mathbf{Z}^d \to \widehat{G} \to G \to 1.
\ee
Concretely, $G$ is called the \emph{point group}. $G$ is not always a group, but rather a quotient $\widehat{G}/\mathbf{Z}^d \simeq G$. Only when the extension is split, will $G$ be a group. For convenience, we refer to $G$ as the point group, even in the nonsymmorphic case. 

The data $H(\mathbf{k})$, $\mathcal{H}(\mathbf{k})$ and $\mathcal{M}$ can be conveniently packaged in terms of a Hilbert bundle: 
\begin{figure}[h]
\centering
\def\svgwidth{2.5cm}
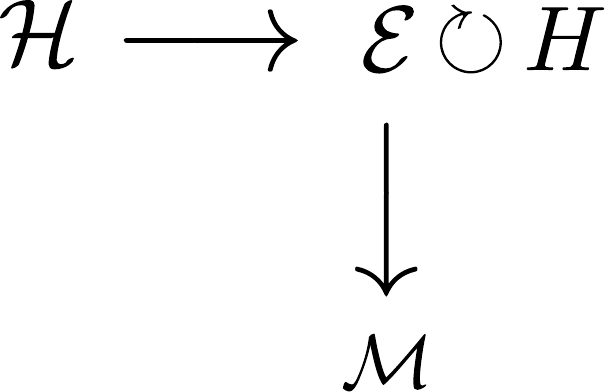
\end{figure}\\
The fibres of this bundle are the Hilbert spaces $\mathcal{H}$, which due to the gapped nature of the system, split into a direct sum $\mathcal{E} = \mathcal{E}_c \oplus \mathcal{E}_v$, with $\mathcal{E}_c$ the conduction band and $\mathcal{E}_v$ the valance band. For an insulator the valance band is completely filled and the fermions, i.e electrons, in those states can have non-trivial behavior. For topological phases, the non-trivial behavior stems from the topology of $\mathcal{E}_v$. For example, the quantization of the Hall conductivity $\s_{xy}$ may seen to be due to the topology of $\mathcal{E}_v$ using the TKNN invariant \cite{Thouless1982, Avron1985}. In fact, the first Chern number of $\mathcal{E}_v$ is directly proportional to $\s_{xy}$. A topologically trivial insulator is then one for which $\mathcal{E}_v$ is topologically trivial. In contrast, for non-trivial topological insulators, the vector bundle $\mathcal{E}_v$ is non-trivial. The different topological types of vector bundles can be enumerated under a suitable notion of equivalence. This enumeration is a classification of vector bundles and thus of topological phases.

\subsection{Briefest introduction to $K$-theory}

The classification of finite rank vector bundles as a mathematical pursuit was initiated in the late 50s and early 60s by Grothendieck and Atiyah with the development of $K$-theory. Since then, this theory has been generalized in several directions. The basic idea of this work can be readily understood by considering a space $X$ which consists of a single point, i.e. $X = \{x\}$. Vector bundles over $x$ are vector spaces of a particular rank $n$. Suppose $V_n$ and $V_{n'}$ are two vector spaces of rank $n$ and $n'$, respectively. In order to classify these vector spaces, we need a notion that compares them. In $K$-theory, the notion of bundle isomorphisms is used. Specifically, in the present example $V_n$ is isomorphic to $V_{n'}$ if and only if $n = n'$. Different vector bundles over $X$ are thus classified by their rank, which is a non-negative integer. Vector bundles can be added using the so-called internal Whitney sum, giving the set of isomorphism classes, $Vect(X)$ the structure of an Abelian monoid. Using the bundle isomorphism, the monoid is isomorphic to $\mathbf{N}$. The resulting set does not form a group (it does not contain inverses), complicating further analysis. Fortunately, however, $Vect(X)$  may be converted to a group using the Grothendieck completion. This construction takes two copies of $Vect(X)$ and subjects it to the following equivalence relation
\bea
(m,n) &\sim& (m',n') \\
\Leftrightarrow \text{there exists $p$ such } &\text{that}& \text{ $m+n'+p=n+m'+p$}.\nonumber
\eea
Let us denote the equivalence classes by $[(m,n)]$. The essential new feature is now that we can take inverses; $[(n,m)]$ is the inverse of $[(m,n)]$. Consequently, elements in the Grothendieck completion are denoted by formal differences, $m-n$. For the case at hand, $\mathbf{N}$ is converted to $\mathbf{Z}$ by the Grothendieck completion. The $K$-theory of $X$ is then said to be $\mathbf{Z}$ and is denoted as $K^0(X) = \mathbf{Z}$. Although we discussed only the zero-dimensional case, for general compact base manifolds $X$, a similar statement has been verified \cite{Karoubi1978} and is encapsulated in the following proposition.
\begin{prop} 
Every element in $K^0(X)$ can be written as $[E]-[\Theta_n]$, where $E$ is a vector bundle over $X$ and $\Theta_n$ is a trivial vector bundle of rank $n$ over $X$. Moreover, $[E]-[\Theta_n] = [F] - [\Theta_p]$ if and only if there exists an integer $q$ such that $E \oplus \Theta_{p+q} \simeq F \oplus \Theta_{n+q}$.
\end{prop} 

From a physics point of view, the proposition is easily interpreted. In section \ref{Appsetup} we discussed how free fermion systems naturally acquire the structure of a Hilbert bundle. The Hilbert bundle $\mathcal{H}$  can have a topology measured for example by the TKNN invariant. Adding a trivial vector bundle $\Theta_m$ to  $\mathcal{H}$ should not change, for example, the conduction properties of the electrons. We thus want to regard $\mathcal{H}$ and $\mathcal{H}' = \mathcal{H} \oplus \Theta_m$ as being topologically equivalent. Indeed, in $K$-theory we see that the trivial piece can be subtracted, i.e. $[\mathcal{H}] = [\mathcal{H}'] - [\Theta_m]$. 

\subsection{Equivariant $K$-theory}
\label{Appequivariant}
\begin{figure}[t]
\centering
\def\svgwidth{6cm}
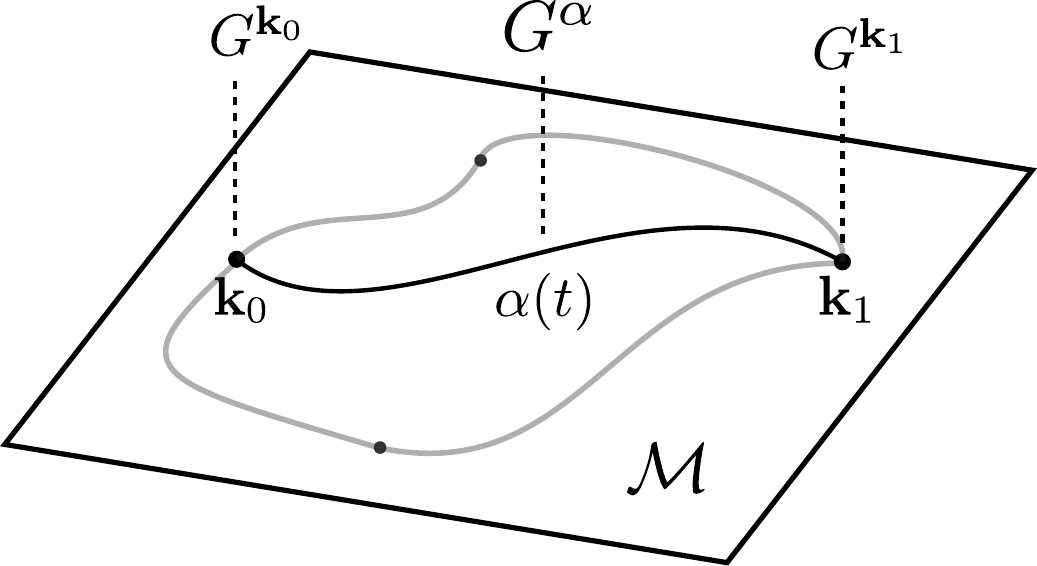
\caption{\label{inducedreps} Representations of $G^{\a}$ along $\a$ induce representations of $G^{\mathbf{k}_i}$ at $\mathbf{k}_i$. The shaded paths indicate possible other fixed point sets in $\mathcal{M}$. }
\end{figure}

The final ingredient in our discussion is the point group symmetry of the lattice. In momentum space, the action of $G$ on $\mathcal{M}$ is defined as
\be\label{geometricaction}
g\cdot \mathbf{k} = D(g)\mathbf{k}
\ee
for $\mathbf{k}$ in $\mathcal{M}$ and $g$ in $G$, the point group. $D(g)$ is a fixed $d$-dimensional representation acting by matrix multiplication on $\mathbf{k}$.
To see that the states in $\mathcal{H}(\mathbf{k})$ form a representation of $G$, it is enough to notice that for $g$ in $G$ the Hamiltonian satisfies, 
\be\label{actionH}
\rho_k(g)H(g\cdot \mathbf{k}) = H(\mathbf{k})\rho_k(g).
\ee
with $\rho_k$ a representation of $G$ under which the states transform. This can be verified using $\rho(g)H = H\rho(g)$ and Equation \eqref{hamsec}. In particular, we will denote the action on the states $\ket{\psi_{\mathbf{k}}}$ in $\mathcal{H}(\mathbf{k})$ by
\be
g*\ket{\psi_{\mathbf{k}}} = \rho_k(g)\ket{\psi_{\mathbf{k}}}
\ee
For generic momenta, the states will form a trivial representation, because Equation \eqref{actionH} is not a commutation relation. Nevertheless, for a subset $S$ of $\mathcal{M}$ which is held pointwise fixed by a subgroup $G^S$, we have $[H(g\cdot \mathbf{k}),\rho_{\mathbf{k}}(g)] = 0$. The states with $\mathbf{k}$ in $S$ can then form non-trivial representations. This extra data provides the Hilbert bundle with an equivariant structure in the sense discussed by Segal in \cite{segal1968equivariant}. In particular, the projection $p: \mathcal{E} \to \mathcal{M}$ is defined as $p(\ket{\psi_{\mathbf{k}}}) = \mathbf{k}$. This implies:
\be
p(g * \ket{\psi_{\mathbf{k}}}) = p(\rho_k(g)\ket{\psi_{g\cdot \mathbf{k}}}) = g\cdot \mathbf{k} = g\cdot p(\ket{\psi_{\mathbf{k}}}),
\ee 
which shows that $p$ respects the action of $G$. Furthermore, the action of $g$ provides a homomorphism between the fibres, i.e. vector spaces, at $k$ and $g\cdot k$. With these properties and the action of $G$ on $\mathcal{M}$ and $\mathcal{E}$, as well as relation \eqref{actionH}, we can choose the representation of the filled states, i.e. $\mathcal{E}_v$,  which is relevant for classifying topological phases in class $A$. Let us make this concrete. Consider a point $\mathbf{k}_0$ in $\mathcal{M}$ and its stabilizer group $G^{\mathbf{k}_0}$ (little co-group in the main text), which leaves $\mathbf{k}_0$ pointwise fixed. The fibre $\mathcal{E}_{\mathbf{k}_0}$ at $\mathbf{k}_0$ is a vector space of eigenstates of $H(\mathbf{k}_0)$. These states form a representation of $G^{\mathbf{k}_0}$ and hence so is $\mathcal{E}_{\mathbf{k}_0}$. Let $\rho_{\mathbf{k}_0}$ be a representation of $G^{\mathbf{k}_0}$. Now consider variations of $\mathbf{k}_0$. We do this by choosing a path $\a(t)$ in $\mathcal{M}$ such that $\a(0) = \mathbf{k}_0$ and $\a(1) = \mathbf{k}_1$ for some $\mathbf{k}_1$ in $\mathcal{M}$, as shown in Figure \ref{inducedreps}.
Along $\a$ a subgroup $G^{\a}$ of $G^{\mathbf{k}}$ is preserved. Without loss of generality we can choose this situation instead of the reverse case with $G^{\mathbf{k}}$ being a subgroup of $G^{\a}$ and we can assume $G^{\a}$ to be independent of $t$. At $\a$, the states form a representation of $G^{\a}$, which we denote by $\widetilde{\rho}_{\a}$. Taking the limit $t \to 0$, $\widetilde{\rho}_{\a}$ induces the representation $\rho_{\mathbf{k}_0}$, meaning that $\rho_{\mathbf{k}_0}|_{G^{\a}} = \rho_{\a}$. This constrains the representations that can be induced at $\mathbf{k}_0$ given a representation along $\a$. Similar arguments hold when taking the limit $t \to 1$. Finding the constraints between all fixed points in this way, results in a finite set of gluing conditions between fibres at different $\mathbf{k}$. These conditions signify the fact that we cannot just pick any number of representation at the fixed points and be guaranteed that we obtain a representation of $\mathcal{E}_v$. 

The constraints can be understood more clearly when considering the representation rings of the various fixed points in $\mathcal{M}$. Consider the example discussed above, with $\mathbf{k}_0$ and $\mathbf{k}_1$ connected by a a path $\a$. The stabilizer groups are $G^{\mathbf{k}_0}, G^{\mathbf{k}_1}$ and $G^{\a}$ respectively. Denote by $R(G^{\mathbf{k}_0})$, $R(G^{\mathbf{k}_1})$, and $R(G^{\a})$ the representation rings over $\mathbf{Z}$ of the stabilizer groups. These rings are constructed by assigning to each irrep of the stabilizer group a copy of $\mathbf{Z}$. We denote the dimensions of the representation rings by $d_0, d_1$ and $d_{\a}$, respectively. The constraints are then maps $\phi_{0,1}$ in
\be
\mathbf{Z}^{d_0} \xrightarrow{\;\phi_0\;} \mathbf{Z}^{d_{\a}} \xleftarrow{\;\phi_1\;} \mathbf{Z}^{d_1},
\ee
which can be represented by the following matrices,
\be
A_{ij}^{\mathbf{k}} = \left\{\begin{array}{cc}
1 & \rho^{i}_{\mathbf{k}}|_{G^{\a}} = \rho^{j}_{\a}\\[0.2cm]
0 & \rho^{i}_{\mathbf{k}}|_{G^{\a}} \neq \rho^{j}_{\a}
\end{array}\right.
\ee
with $\mathbf{k} = \mathbf{k}_{0,1}$. 

The task of finding the constraints between all fixed points can be simplified using the fact that $g : \mathcal{E}_k \to \mathcal{E}_{g\cdot k}$ is a homomorphism. We can use this homomorphism to focus on just the fundamental domain $\W$ of the action of $G$. This domain only includes points $k$ that are not related to each other by an element of $G$, and hence knowing the constraints in that region is enough to know all constraints in all of $\mathcal{M}$. 

As an example, consider the $p4mm$ case, which was detailed in the main text. Let us concentrate on $\a = l_1$ with $k_0 = \G$ and $k_1 = X$. In that case, the matrices $A^{\mathbf{k}}$ take the form
\be
A^{\G} = \begin{pmatrix}
1 & 0 & 1 & 0 & 1\\
0 & 1 & 0 & 1 & 1
\end{pmatrix},\quad
A^{X} = \begin{pmatrix}
1 & 0 & 1 & 0\\
0 & 1 & 0 & 1
\end{pmatrix}.
\ee
In fact, since both images of $A^{\G}$ and $A^{X}$ are $\mathbf{Z}^{l_1} = \mathbf{Z}^2$, we can combine the above relations in two relations. Denoting elements in a representation ring by $\mathbf{n}^{\mathbf{k}} = (n_0^{\mathbf{k}}, \dots, n_{d_{\mathbf{k}}}^{\mathbf{k}})^T$, we have
\be
A^{\G}\mathbf{n}^{\G} = A^X\mathbf{n}^{X}
\ee

Imposing these relations on the elements in the representation rings at each fixed point in $\W$, gives us a set of independent integers that specify the representation of $\mathcal{E}$ and in particular of $\mathcal{E}_v$. In the above, we did not specify the type of representation, and thus the same arguments also hold for projective representations. These representations occur when the group extension in \eqref{groupextension} is not split, i.e. for nonsymmorphic crystals. The integers discussed above thus fix the representation of $\mathcal{E}_v$ for both split and non-split extensions in \eqref{groupextension}. However, they do not fix the characteristic classes of $\mathcal{E}_v$. For the complex vector bundles discussed above, the Chern character is the most important one and results in an integer once integrated over an even dimensional submanifold of the base manifold $\mathcal{M}$. Below in section \ref{AppChern} we discuss how these are constrained by space group symmetry.

In summary, to fix the topology and representation of the $G$-equivariant bundle $\mathcal{E} \to \mathcal{M}$, one specifies the Chern numbers and the set of independent integers in each representation ring associated to each fixed point. 

The point of view we have taken in the above discussion is in fact an easy way of understanding the classification of $G$-equivariant vector bundles. The integers and Chern numbers discussed there are the only integers that need to be specified to fix an equivalence class in $K$-theory. In fact, the $K$-theory of $\mathcal{M}$ given an action of $G$ computes Abelian invariants of $G$-equivariant bundles over $\mathcal{M}$. These are the representation of the bundle and the Chern numbers. The $K$-theory is $K^0_{G}(\mathcal{M})$. In terms of cyrstal symmetries, this is true for both symmorphic and nonsymmorphic crystals. In conclusion, topological phases in class $A$ protected by space group $\widehat{G}$ are classified by 
\be
\text{TopPh}_{\widehat{G}}^d = K^{0}_{G}(\mathbf{T}^d).
\ee
A similar conclusion was found in \cite{Freed2013}. For class $AIII$ we can use the same type of $K$-theory, but now $K^0$ is replaced by $K^{-1}$.  

From a mathematical point of view this in principle concludes the analysis, were it not for the fact that these $K$-theories are in general very hard to compute. Although, for a given a space group $\widehat{G}$, mathematical precise methods are being developed to compute $K_{G}^0(\mathbf{T}^d)$ \cite{RoyerFreed}, the arguments and results of the present work provide an intuitive and simple way of performing that computation.

\subsubsection{Relations between $K$-theory and de Rham cohomology}

Before going into actual examples, we briefly discuss some interesting results relating $K$-theory to de Rham cohomology. These results help us to check the arguments we made above and in the main text. Details can be found in \cite{Karoubi1978, Park2008, Hirzebruch1990}. 

To make the connection with de Rham cohomology we use the Chern character. The Chern character assigns an even dimensional form to a vector bundle $\mathcal{E}$ as
\be
\text{Ch}(\mathcal{E}) = \sum_{n=0}^{\infty}\frac{1}{n!}\Tr\left( \frac{i\mathcal{F}_{\mathcal{E}}}{2\pi} \right)^n
\ee  
where $\mathcal{F}_{\mathcal{E}}$ is the Berry curvature two-form of the bundle $\mathcal{E}$. The $n$-th Chern character is given by
\be\label{cherncharac}
\text{Ch}_n(\mathcal{E}) = \frac{1}{n!}\Tr\left( \frac{i\mathcal{F}_{\mathcal{E}}}{2\pi} \right)^n.
\ee
From the $K$-theory perspective, the Chern character provides a (ring) homomorphism
\bea
\text{Ch} : K^0(\mathcal{M}) &\to & H^{\text{even}}_{dR}(\mathcal{M};\mathbf{Q}) = \bigoplus_{n=0}^{\infty} H^{2n}_{dR}(\mathcal{M};\mathbf{Q}),\nonumber\\ \text{Ch}&:&[E - \Theta_m]\mapsto \text{Ch}([E])
\eea
where $H^{2j}_{dR}(\mathcal{M};\mathbf{Q})$ is the $2j$-th de Rahm cohomology of $\mathcal{M}$ over the rational $\mathbf{Q}$. Forgetting about torsion in $K^0(\mathcal{M})$, this becomes an isomorphism $K^0(\mathcal{M}) \otimes \mathbf{Q} \simeq H^{\text{even}}_{dR}(\mathcal{M})$. For the $K$-theory, $K^{-1}(\mathcal{M})$, a similar statement holds. In fact,
\be
\text{Ch} : K^{-1}(\mathcal{M}) \to H^{\text{odd}}_{dR}(\mathcal{M};\mathbf{Q})
\ee
is a group homomorphism and 
\be
H^{\text{odd}}_{dR}(\mathcal{M};\mathbf{Q}) = \bigoplus_{n=0}^{\infty} H^{2n+1}_{dR}(\mathcal{M};\mathbf{Q}).
\ee

These odd-dimensional cohomology classes are generated by odd-dimensional forms of the form $\Tr((f^{-1}df)^{2n+1}))$ with $f: \mathcal{M} \to \mathcal{E}$ a smooth function on $\mathcal{M}$. In contrast to the Chern characters, they can be understood as winding numbers once integrated. A similar isomorphism as for $K^0(\mathcal{M})$ also exists in this case:
\be
K^*(\mathcal{M}) \otimes \mathbf{Q} \simeq H^*_{dR}(\mathcal{M};\mathbf{Q}).
\ee
The purpose of this isomorphism is to translate information hidden in the $K$-theory of $\mathcal{M}$ to a more familiar form in terms of the cohomology of $\mathcal{M}$. The Chern characters in \eqref{cherncharac} give an accurate account for this information, which is most easily seen by integrating them. In doing so, the characters $\text{Ch}_n$ become topological invaraints, called Chern numbers $c_n$. These Chern numbers can only take integer values and account for various topological properties of gapped free fermion systems. An example of a physical observable related to Chern numbers is the TKNN invariant. This invariant is directly proportional to the first Chern number and is related to the Hall conductivity by 
\be
\s_{xy} = \frac{e^2}{h} \int_{\mathbf{T}^2} \text{Ch}_1(\mathcal{E}) = \frac{e^2}{h}c_1.
\ee
Here the integration is over the full Brillouin zone $\mathcal{M} = \mathbf{T}^2$. The winding numbers obtained from integrating odd forms in $H^{\text{odd}}_{dR}$ also capture information about the topological phase, but then for those in class $AIII$. We note that they are related to electric and magnetoelectric polarizability \cite{Qi2008}, but will not discuss them here.
 
In the present context of electrons within a crystal lattice, we will mostly be interested in $\mathcal{M} = \mathbf{T}^2$ or $\mathbf{T}^3$, and hence we will only be concerned with the zeroth and first Chern number. In the equivariant picture sketched above, we already saw how the zeroth Chern number, which is responsible for the representations, is constrained by space group symmetry. We will now see how these contrains affect the first Chern number. 

\subsubsection{Chern numbers and crystal symmetry}
\label{AppChern}
Chern numbers can only be defined on even dimensional fixed submanifolds of $\mathbf{T}^3$. In the case at hand, these are the bounding planes $P$ in $\W$. These are planes in the fundamental domain, but to integrate the Chern character, we need a submanifold $\mathcal{N}$ in $\mathbf{T}^3$. Denote by $C_g$ the centralizer of the symmetry $g$ that leaves $P$ invariant. This submanifold $\mathcal{N}$ is then obtained by acting with $C_g$ on $P$. In other words, $C_g$ still has a non-trivial action on the submanifold. When the Chern characters are integrated, the action of $C_g$ needs to be taken into account. The action of $C_g$ can be such that it inverts the orientation of $\mathcal{N}$. Thus when unfolding $P$, $\mathcal{N}$ will consist of patches $U_-$ and $U_+$ with different orientations, indicated by the subscript. More precisely, the submanifold $\mathcal{N}$ is a $|C_g|$-fold cover of $P$ and whenever orientation reversing elements are in $C_g$, $|C_g|$ is even. Suppose $g^*$ is the orientation reversing element. The submanifold $\mathcal{N}$ will then contain an equal number of patches $U_+$ and $U_- = g^* U_+$. The integral of the Chern character over $\mathcal{N}$ will therefore vanish. Thus, whenever there are orientation reversing elements in $C_g$, the Chern numbers on planes fixed by $g$ will be zero. 

To be a bit more explicit, consider $\mathcal{N}$ to be two dimensional, i.e. $\mathcal{N} \simeq \mathbf{T}^2$. This two-torus is held fixed by $G^{\mathcal{N}} \simeq \mathbf{Z}_2 = \braket{g}$ and has centralizer $C_g$. In the fundamental domain we denote it as the subset $P$, thus $\mathcal{N}$ is a $|C_g|$-fold cover of $P$. For topological phases in real materials, this is the only case of interest. Let $\mathcal{F}_{\mathcal{E}} = F_{xy}\;dk_x \wedge dk_y$ be the Berry curvature of a vector bundle $\mathcal{E}$. Suppose $h$ is an element of $C_g$, then $h$ acts on $\mathcal{F}_{\mathcal{E}}$ as
\be
F_{xy}(D(h)\mathbf{k}) = \det(D(h)) F_{xy}(\mathbf{k}) 
\ee
with $D$ a fixed representation, as in equation \eqref{geometricaction}. The Chern number is given by
\begin{eqnarray}\label{chern}
c_1 &=& \int_{\mathcal{N}} F_{xy}(k_x,k_y)\;d^2k  \\&=& \sum_{h\in C_g} \det(D(h))\int_{P} F_{xy}(k_x,k_y)\;d^2k \nonumber.
\end{eqnarray}
The sum will tell us whether $c_1$ vanishes or not. The centralizer can either be $\mathbf{Z}_n$ or $D_n$ with $n = 2$, $3$, $4$, or $6$, or it can be trivial. When it is one of the cyclic groups, then $c_1$ does not vanish, but when $C_g$ contains reflections, half of the terms in the sum in \eqref{chern} have negative determinant, ensuring that  $c_1 = 0$. 

\subsubsection{Segal's formula}

The maps relating $K$-theory to ordinary cohomology are useful when considering the following result by Segal \cite{Hirzebruch1990},
\be\label{segal}
K^{-n}_G(\mathcal{M}) \otimes \mathbf{C} = \bigoplus_{[g]} K^{-n}(\mathcal{M}^g)^{C_g}\otimes \mathbf{C}.
\ee
with $\mathcal{M}$ compact and $G$ finite. The sum is over representatives of conjugacy classes of $G$. The centralizer of $g$ is denoted by $C_g$ and $\mathcal{M}^g$ is the fixed point set of $g$. This formula relates $G$-equivariant $K$-theory (tensored with $\mathbf{C}$) to the ordinary $K$-theory of the fixed points $\mathcal{M}^g$. To deal with the $C_g$ acting on $K^{-n}(\mathcal{M}^g)$ we compose this formula with the Chern character $\text{Ch}$ as discussed above. The summand on the right hand side then amounts to counting invariant forms on $\mathcal{M}^g$.

\subsection{Example: Octahedral group}

To see how all this applies to a non-trivial example, consider again the case that was also discussed in the main text; $G = \mathbf{Z}_2 \times S_4$. The corresponding group represents the symmetries of the cube and is generated by three elements, which act on $\mathcal{M} = \mathbf{T}^3$ as
\bea\label{Ohaction}
r\cdot k = (-k_z,k_y,k_x)&,& \quad I \cdot k = -k,\\\nonumber t\cdot k &=& (k_y, k_x, -k_z).
\eea
\begin{figure}[t]
\centering
\def\svgwidth{4cm}
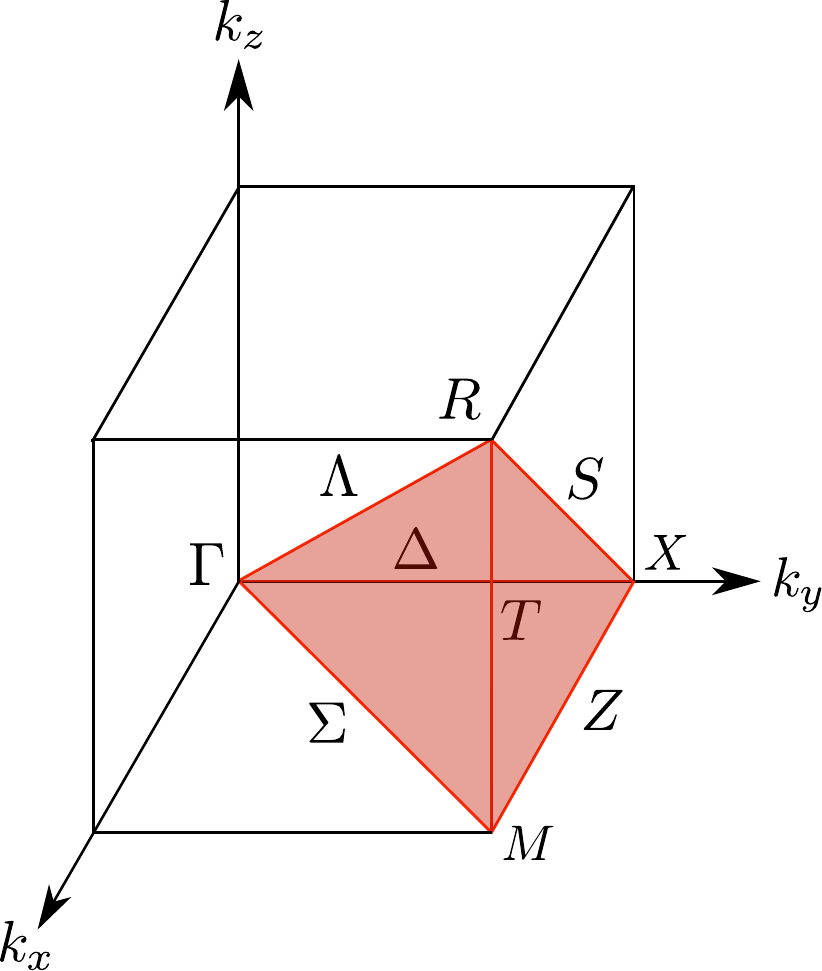
\caption{\label{FundOh2} Fundamental domain $\W$ of the octahedral group $\mathbf{Z}_2 \times S_4$ (shaded red). The red lines are the high symmetry lines $\Lambda$, $\Delta$, $\Sigma$, $S$, $T$, and $Z$}
\end{figure}
\noindent 
The space group containing this point group that we want to consider is $Pm\bar{3}m$. This space group is symmorphic, meaning that 
\be
1 \to \mathbf{Z}^3 \to Pm\bar{3}m \to \mathbf{Z}_2 \times S_4 \to 1.
\ee
is split and hence we only need to compute ordinary irreducible representations at the fixed points. From Fig. \ref{FundOh2} we see that the fixed points are $\G$, $R$, $X$ and $M$. The stabilizer groups of the fixed points are $G^{\G} = G^R = \mathbf{Z}_2 \times S_4$ and $G^M = G^X = \mathbf{Z}_2 \times D_4$. The fixed lines are $G^Z = G^{\Sigma} = G^S = \mathbf{Z}_2 \times \mathbf{Z}_2$, $G^{\Delta} = G^T = D_4$ and $G^{\Lambda} = S_3$. The bounding planes of the fundamental domain $\W$ are held fixed by a $\mathbf{Z}_2$ subgroup of $\mathbf{Z}_2 \times S_4$, which acts as a reflection. 

At the fixed points, the representation rings are $R(G^{\G}) = R(G^{R}) = \mathbf{Z}^{10}$, $R(G^{M}) = R(G^{X}) = \mathbf{Z}^{10}$. The six fixed lines in the fundamental domain result in $25$ relations between the $40$ integers coming from these rings. This results in only $22$ integers that are independent and specify the representation of a $G$-equivariant vector bundle over $\mathbf{T}^3$. Moreover,  the centralizers of fixed planes always contain a reflection and hence all Chern numbers vanish. From this we can conclude that:
\be\label{resultOh}
K^0_{\mathbf{Z}_2 \times S_4}(\mathbf{T}^3) \simeq \mathbf{Z}^{22}.
\ee
Let us now show that this computation agrees with equation \eqref{segal}, proposed by Segal. In this formula, we need to compute the fixed point manifolds associated with representatives of the conjugacy classes of $G$. This information is collected in table \ref{tableOhsegal}.
\begin{table}
\centering
\begin{tabular}{lll}
$[g]$ & $\mathcal{M}^g$ & $C_g$\\\hline
$[1]$ & $\mathbf{T}^3$ & $G$ \\
$[tr^2tr^2]$ & $\{(0,0,k_z), (\pi,\pi,k_z), (0, \pi, k_z),$ & $\mathbf{Z}_2 \times D_4$ \\
& $\;\;(\pi, 0, k_z)\}$ & \\
$[r^3tr]$ & $\{(0,k_y,-k_y), (\pi,k_y,-k_y)\}$ & $\mathbf{Z}_2^3$\\
$[Itrt]$ & $\{ (0,0,0), (0,\pi,\pi), (\pi,0,0),$ & $\mathbf{Z}_2 \times \mathbf{Z}_4$\\
& $\;\;(\pi,\pi,\pi) \}$ & \\
$[Itr^2t]$ & $\{(0,k_y,k_z),(\pi,k_y,k_z)\}$ & $\mathbf{Z}_2 \times D_4$\\
$[tI]$ & $\{(k_x,-k_x,k_z)\}$ & $\mathbf{Z}_2^3$\\
$[Itr^3]$ & $\{(0,0,0),(\pi,\pi,\pi)\}$ & $\mathbf{Z}_6$ \\
$[rtr^3t]$ & $\{(k_x,-k_x,k_x)\}$ & $\mathbf{Z}_6$\\
$[r^2t]$ & $\{(0,0,k_z),(\pi,\pi,k_z)\}$ & $\mathbf{Z}_2 \times \mathbf{Z}_4$\\
$[I]$ & $\{\; 8 \; \text{pnts}\}$ & $G$
\end{tabular}
\caption{\label{tableOhsegal} Fixed point sets $\mathcal{M}^g$ for each conjugacy class $[g]$ of $G$ and their centralizers $C_g$. The set $\{ 8  \text{ pnts}\}$ equals $\{ (0,0,0), (\pi,0,0), (0,\pi,0), (0,0,\pi), (\pi,\pi,0), (\pi,0,\pi),$ $(0,\pi,\pi), (\pi,\pi,\pi) \}$.} 
\end{table}
From this table, the integers found in \eqref{resultOh} can easily be extracted. The only thing that we need to take into account, is the action of $C_g$ on $\mathcal{M}^g$. Let us start with the Chern numbers. For $[1]$, $[Itr^2t]$ and $[tI]$ no Chern numbers are possible, because the centralizer contains a reflection. The $K$-theory is then,
\bea
&K&^0(\mathbf{T}^3)^G \otimes \mathbf{C} \oplus K^0(\mathbf{T}^2 \amalg \mathbf{T}^2)^{\mathbf{Z}_2\times D_4} \otimes\nonumber \\ &\mathbf{C}& \oplus K^0(\mathbf{T}^2)^{\mathbf{Z}_2^3} \otimes \mathbf{C} \simeq \mathbf{C}^4
\eea
Similarly, for the one-dimensional fixed point sets we get eight copies of $\mathbf{C}$, because for $[tr^2tr^2]$ two circles are related by the action of $C_{tr^2tr^2} = \mathbf{Z}_2 \times D_4$. Finally, the zero-dimensional fixed point sets give $\mathbf{C}^{10}$ following the same reasoning as before. Adding these results all up, we obtain
\be\label{resultsegalOh}
K^0_G(\mathbf{T}^3)\otimes \mathbf{C} \simeq \mathbf{C}^{22}.
\ee
Hence, up to torsion, equations \eqref{resultOh} and \eqref{resultsegalOh} exactly agree, as anticipated. Fortunately, for class $A$ there is no torsion present for both two and three dimensions \cite{RoyerFreed} and so our method gives the correct result for $K^0_{\mathbf{Z}_2\times S_4}(\mathbf{T}^3)$. It also confirms that our method gives the correct results both for symmorphic and nonsymmorphic two dimensional space groups. Although we have not check all 230 space groups in three dimensions, we believe that our method does still agree with the exact $K$-theory computation. 
\bibliographystyle{apsrev4-1}
\bibliography{CSpaper}
\end{document}

%% file: FundDomainp4mm.pdf_tex
%% Creator: Inkscape inkscape 0.91, www.inkscape.org
%% Accompanies image file 'FundDomainp4mm.pdf' (pdf, eps, ps)
\begingroup%
  \makeatletter%
  \providecommand\color[2][]{%
    \errmessage{(Inkscape) Color is used for the text in Inkscape, but the package 'color.sty' is not loaded}%
    \renewcommand\color[2][]{}%
  }%
  \providecommand\transparent[1]{%
    \errmessage{(Inkscape) Transparency is used (non-zero) for the text in Inkscape, but the package 'transparent.sty' is not loaded}%
    \renewcommand\transparent[1]{}%
  }%
  \providecommand\rotatebox[2]{#2}%
  \ifx\svgwidth\undefined%
    \setlength{\unitlength}{275.20561639bp}%
    \ifx\svgscale\undefined%
      \relax%
    \else%
      \setlength{\unitlength}{\unitlength * \real{\svgscale}}%
    \fi%
  \else%
    \setlength{\unitlength}{\svgwidth}%
  \fi%
  \global\let\svgwidth\undefined%
  \global\let\svgscale\undefined%
  \makeatother%
  \begin{picture}(1,0.9203596)%
    \put(0,0){\includegraphics[width=\unitlength,page=1]{FundDomainp4mm.pdf}}%
  \end{picture}%
\endgroup%

%% file: WeylConeS+D.pdf_tex
%% Creator: Inkscape inkscape 0.91, www.inkscape.org
%% Accompanies image file 'WeylConeS+D.pdf' (pdf, eps, ps)
\begingroup%
  \makeatletter%
  \providecommand\color[2][]{%
    \errmessage{(Inkscape) Color is used for the text in Inkscape, but the package 'color.sty' is not loaded}%
    \renewcommand\color[2][]{}%
  }%
  \providecommand\transparent[1]{%
    \errmessage{(Inkscape) Transparency is used (non-zero) for the text in Inkscape, but the package 'transparent.sty' is not loaded}%
    \renewcommand\transparent[1]{}%
  }%
  \providecommand\rotatebox[2]{#2}%
  \ifx\svgwidth\undefined%
    \setlength{\unitlength}{568.28884335bp}%
    \ifx\svgscale\undefined%
      \relax%
    \else%
      \setlength{\unitlength}{\unitlength * \real{\svgscale}}%
    \fi%
  \else%
    \setlength{\unitlength}{\svgwidth}%
  \fi%
  \global\let\svgwidth\undefined%
  \global\let\svgscale\undefined%
  \makeatother%
  \begin{picture}(1,0.52841482)%
    \put(0,0){\includegraphics[width=\unitlength,page=1]{WeylConeS+D.pdf}}%
  \end{picture}%
\endgroup%

%% file: WeylAtHighPnt.pdf_tex
%% Creator: Inkscape inkscape 0.91, www.inkscape.org
%% Accompanies image file 'WeylAtHighPnt.pdf' (pdf, eps, ps)
\begingroup%
  \makeatletter%
  \providecommand\color[2][]{%
    \errmessage{(Inkscape) Color is used for the text in Inkscape, but the package 'color.sty' is not loaded}%
    \renewcommand\color[2][]{}%
  }%
  \providecommand\transparent[1]{%
    \errmessage{(Inkscape) Transparency is used (non-zero) for the text in Inkscape, but the package 'transparent.sty' is not loaded}%
    \renewcommand\transparent[1]{}%
  }%
  \providecommand\rotatebox[2]{#2}%
  \ifx\svgwidth\undefined%
    \setlength{\unitlength}{489.35484398bp}%
    \ifx\svgscale\undefined%
      \relax%
    \else%
      \setlength{\unitlength}{\unitlength * \real{\svgscale}}%
    \fi%
  \else%
    \setlength{\unitlength}{\svgwidth}%
  \fi%
  \global\let\svgwidth\undefined%
  \global\let\svgscale\undefined%
  \makeatother%
  \begin{picture}(1,1.1514718)%
    \put(0,0){\includegraphics[width=\unitlength,page=1]{WeylAtHighPnt.pdf}}%
  \end{picture}%
\endgroup%

%% file: DirectTransition.pdf_tex
%% Creator: Inkscape inkscape 0.91, www.inkscape.org
%% Accompanies image file 'DirectTransition.pdf' (pdf, eps, ps)
\begingroup%
  \makeatletter%
  \providecommand\color[2][]{%
    \errmessage{(Inkscape) Color is used for the text in Inkscape, but the package 'color.sty' is not loaded}%
    \renewcommand\color[2][]{}%
  }%
  \providecommand\transparent[1]{%
    \errmessage{(Inkscape) Transparency is used (non-zero) for the text in Inkscape, but the package 'transparent.sty' is not loaded}%
    \renewcommand\transparent[1]{}%
  }%
  \providecommand\rotatebox[2]{#2}%
  \ifx\svgwidth\undefined%
    \setlength{\unitlength}{488.45733883bp}%
    \ifx\svgscale\undefined%
      \relax%
    \else%
      \setlength{\unitlength}{\unitlength * \real{\svgscale}}%
    \fi%
  \else%
    \setlength{\unitlength}{\svgwidth}%
  \fi%
  \global\let\svgwidth\undefined%
  \global\let\svgscale\undefined%
  \makeatother%
  \begin{picture}(1,1.3426819)%
    \put(0,0){\includegraphics[width=\unitlength,page=1]{DirectTransition.pdf}}%
  \end{picture}%
\endgroup%

%% file: compatibility21.pdf_tex
%% Creator: Inkscape inkscape 0.91, www.inkscape.org
%% Accompanies image file 'compatibility21.pdf' (pdf, eps, ps)
\begingroup%
  \makeatletter%
  \providecommand\color[2][]{%
    \errmessage{(Inkscape) Color is used for the text in Inkscape, but the package 'color.sty' is not loaded}%
    \renewcommand\color[2][]{}%
  }%
  \providecommand\transparent[1]{%
    \errmessage{(Inkscape) Transparency is used (non-zero) for the text in Inkscape, but the package 'transparent.sty' is not loaded}%
    \renewcommand\transparent[1]{}%
  }%
  \providecommand\rotatebox[2]{#2}%
  \ifx\svgwidth\undefined%
    \setlength{\unitlength}{223.69433546bp}%
    \ifx\svgscale\undefined%
      \relax%
    \else%
      \setlength{\unitlength}{\unitlength * \real{\svgscale}}%
    \fi%
  \else%
    \setlength{\unitlength}{\svgwidth}%
  \fi%
  \global\let\svgwidth\undefined%
  \global\let\svgscale\undefined%
  \makeatother%
  \begin{picture}(1,0.97191251)%
    \put(0,0){\includegraphics[width=\unitlength,page=1]{compatibility21.pdf}}%
  \end{picture}%
\endgroup%

%% file: FundDomainOh.pdf_tex
%% Creator: Inkscape inkscape 0.91, www.inkscape.org
%% Accompanies image file 'FundDomainOh.pdf' (pdf, eps, ps)
\begingroup%
  \makeatletter%
  \providecommand\color[2][]{%
    \errmessage{(Inkscape) Color is used for the text in Inkscape, but the package 'color.sty' is not loaded}%
    \renewcommand\color[2][]{}%
  }%
  \providecommand\transparent[1]{%
    \errmessage{(Inkscape) Transparency is used (non-zero) for the text in Inkscape, but the package 'transparent.sty' is not loaded}%
    \renewcommand\transparent[1]{}%
  }%
  \providecommand\rotatebox[2]{#2}%
  \ifx\svgwidth\undefined%
    \setlength{\unitlength}{236.54118095bp}%
    \ifx\svgscale\undefined%
      \relax%
    \else%
      \setlength{\unitlength}{\unitlength * \real{\svgscale}}%
    \fi%
  \else%
    \setlength{\unitlength}{\svgwidth}%
  \fi%
  \global\let\svgwidth\undefined%
  \global\let\svgscale\undefined%
  \makeatother%
  \begin{picture}(1,1.18116989)%
    \put(0,0){\includegraphics[width=\unitlength,page=1]{FundDomainOh.pdf}}%
  \end{picture}%
\endgroup%

%% file: Hilbertbundle.pdf_tex
%% Creator: Inkscape inkscape 0.91, www.inkscape.org
%% Accompanies image file 'Hilbertbundle.pdf' (pdf, eps, ps)
\begingroup%
  \makeatletter%
  \providecommand\color[2][]{%
    \errmessage{(Inkscape) Color is used for the text in Inkscape, but the package 'color.sty' is not loaded}%
    \renewcommand\color[2][]{}%
  }%
  \providecommand\transparent[1]{%
    \errmessage{(Inkscape) Transparency is used (non-zero) for the text in Inkscape, but the package 'transparent.sty' is not loaded}%
    \renewcommand\transparent[1]{}%
  }%
  \providecommand\rotatebox[2]{#2}%
  \ifx\svgwidth\undefined%
    \setlength{\unitlength}{173.99002436bp}%
    \ifx\svgscale\undefined%
      \relax%
    \else%
      \setlength{\unitlength}{\unitlength * \real{\svgscale}}%
    \fi%
  \else%
    \setlength{\unitlength}{\svgwidth}%
  \fi%
  \global\let\svgwidth\undefined%
  \global\let\svgscale\undefined%
  \makeatother%
  \begin{picture}(1,0.64794066)%
    \put(0,0){\includegraphics[width=\unitlength,page=1]{Hilbertbundle.pdf}}%
  \end{picture}%
\endgroup%

%% file: inducedreps.pdf_tex
%% Creator: Inkscape inkscape 0.91, www.inkscape.org
%% Accompanies image file 'inducedreps.pdf' (pdf, eps, ps)
\begingroup%
  \makeatletter%
  \providecommand\color[2][]{%
    \errmessage{(Inkscape) Color is used for the text in Inkscape, but the package 'color.sty' is not loaded}%
    \renewcommand\color[2][]{}%
  }%
  \providecommand\transparent[1]{%
    \errmessage{(Inkscape) Transparency is used (non-zero) for the text in Inkscape, but the package 'transparent.sty' is not loaded}%
    \renewcommand\transparent[1]{}%
  }%
  \providecommand\rotatebox[2]{#2}%
  \ifx\svgwidth\undefined%
    \setlength{\unitlength}{298.63772415bp}%
    \ifx\svgscale\undefined%
      \relax%
    \else%
      \setlength{\unitlength}{\unitlength * \real{\svgscale}}%
    \fi%
  \else%
    \setlength{\unitlength}{\svgwidth}%
  \fi%
  \global\let\svgwidth\undefined%
  \global\let\svgscale\undefined%
  \makeatother%
  \begin{picture}(1,0.54529841)%
    \put(0,0){\includegraphics[width=\unitlength,page=1]{inducedreps.pdf}}%
  \end{picture}%
\endgroup%

%% file: FundDomainOh2.pdf_tex
%% Creator: Inkscape inkscape 0.91, www.inkscape.org
%% Accompanies image file 'FundDomainOh2.pdf' (pdf, eps, ps)
\begingroup%
  \makeatletter%
  \providecommand\color[2][]{%
    \errmessage{(Inkscape) Color is used for the text in Inkscape, but the package 'color.sty' is not loaded}%
    \renewcommand\color[2][]{}%
  }%
  \providecommand\transparent[1]{%
    \errmessage{(Inkscape) Transparency is used (non-zero) for the text in Inkscape, but the package 'transparent.sty' is not loaded}%
    \renewcommand\transparent[1]{}%
  }%
  \providecommand\rotatebox[2]{#2}%
  \ifx\svgwidth\undefined%
    \setlength{\unitlength}{236.54118095bp}%
    \ifx\svgscale\undefined%
      \relax%
    \else%
      \setlength{\unitlength}{\unitlength * \real{\svgscale}}%
    \fi%
  \else%
    \setlength{\unitlength}{\svgwidth}%
  \fi%
  \global\let\svgwidth\undefined%
  \global\let\svgscale\undefined%
  \makeatother%
  \begin{picture}(1,1.18116989)%
    \put(0,0){\includegraphics[width=\unitlength,page=1]{FundDomainOh2.pdf}}%
  \end{picture}%
\endgroup%